\documentclass[11pt]{article}
\usepackage{emulateapj}
\usepackage{lscape}

\newcommand{\HI}{H$\;${\small\rm I}\relax}
\newcommand{\HII}{H$\;${\small\rm II}\relax}
\newcommand{\Ha}{H$\alpha$\relax}

\newcommand{\z}{$z$}
\newcommand{\msun}{M$_\odot$}

\newcommand{\etal}{{et al.}}
\newcommand{\column}{cm$^{-2}$}
\newcommand{\percc}{cm$^{-3}$}

\newcommand{\e}[1]{10^{#1}}
\newcommand{\kms}{km s$^{-1}$}

\newcommand{\garciaburillo}{Garc\'{\i}a-Burillo}
\newcommand{\idfigs}{\ref{fig:IDmid}, \ref{fig:IDne}, 
and \ref{fig:IDsw}}

\newcommand{\hst}{{\em HST}}

\newcommand{\pI}{Paper~I}
\newcommand{\pII}{Paper~II}
\newcommand{\ngc}{NGC~}

\righthead{Multiphase Halo of NGC~891}
\lefthead{Howk \& Savage}

\begin{document}
\submitted{Accepted for publication in {\em The Astronomical
Journal}}

\title{The Multiphase Halo of NGC~891: WIYN \Ha\ and BVI 
Imaging\altaffilmark{1}}

\altaffiltext{1}{Based on observations obtained at the WIYN
Observatory, a joint facility of the University of Wisconsin-Madison,
Indiana University, Yale University, and the National Optical
Astronomy Observatories.}

\author{J. Christopher Howk\altaffilmark{2} \& Blair D. Savage}
\affil{Department of Astronomy, University of Wisconsin-Madison, 
        Madison, WI 53706 \\ Electronic mail:
     howk@astro.wisc.edu, savage@astro.wisc.edu}

\altaffiltext{2}{Current address: Dept. of Physics \& Astronomy, The
Johns Hopkins University, 3400 N. Charles St., Baltimore, MD, 21218;
e-mail:~howk@pha.jhu.edu.}

\authoremail{howk@pha.jhu.edu}


\begin{abstract}
We present new, deep optical images (BVI+\Ha) of the interstellar
medium (ISM) far above the plane of the edge-on Sb galaxy \ngc 891.
These sub-arcsecond ($0\farcs8 - 0\farcs9$) ground-based images give a
direct visual view of two physically distinct ``phases'' of the thick
interstellar disk of this galaxy.  A dense, likely cold, phase of the
thick disk ISM is observed in our BVI images as highly-structured
dust-bearing clouds viewed in absorption against the background
stellar light of the galaxy.  These dusty structures are traceable to
heights $|z| \sim 2$ kpc from the midplane.  Though our data are deep
enough to identify such clouds at higher \z, very few
highly-structured dust features are present at $|z| \ga 2$ kpc.  If
the more prominent dust structures have Galactic gas-to-dust ratios,
then they have gas masses in excess of $10^5$ \msun, each having
visual extinctions well in excess of unity.  A warm ionized phase of
the high-\z\ ISM is observed through its well-studied \Ha\ emission.
Our images of this medium, to date the highest-resolution observations
of the warm ionized medium in \ngc 891, show that the diffuse ionized
medium in this galaxy is relatively smoothly distributed with some
filamentary structure superposed on this smooth background.  Our data
show very little correspondence between the \Ha -emitting material and
the absorbing dust structures.  {\em These two phases of the
multiphase high-z ISM are physically distinct.}  The material traced
by \Ha\ emission is much more smoothly distributed than that traced by
the absorbing dust clouds, and it is clear that the \Ha\ emission is
being heavily extincted in many places by the dense dust-bearing
medium.  The high-\z\ dust clouds in \ngc 891 are a major component of
the multiphase halo medium.  The mass found in the cold dense, warm
neutral, and warm ionized phases are for $|z| > 0.4$ kpc are
comparable and of order a few$\times\e{8}$ \msun.  Our discussion of
the physics of the multiphase medium at high-\z\ relies heavily on the
theoretical work of Wolfire \etal, and we suggest the dense phase of
the thick disk ISM exists to $|z| \la 2$ kpc because the thermal
pressures of the warm and hot phases in this region are sufficient for
its maintenance.  Our \Ha\ observations show evidence for several
discrete \HII\ regions at large distances from the midplane ($0.6 \la
|z| \la 2.0$ kpc).  The presence of these \HII\ regions in the thick
disk of \ngc 891 suggests that on-going star formation may be present
in some of the dense, high-\z\ clouds visible in our images.

\end{abstract}


\section{Introduction}

The interstellar medium (ISM) in the disk of the Milky Way (and other
spirals) is made up of several thermal ``phases,'' each with a
characteristic density and temperature depending on the prevailing
physical conditions (Field, Goldsmith, \& Habing 1969; McKee \&
Ostriker 1977).  The principal phases of the neutral material in the
disks of spiral galaxies, the cold neutral medium (CNM) and the warm
neutral medium (WNM), are expected to exist in rough pressure
equilibrium, embedded in the warm and hot ionized phases of the ISM.
Under the right conditions, the CNM can include a molecular component
from which young stars form.

The conditions in the equilibrium neutral media are dictated by the
ambient pressure and the detailed heating and cooling processes in the
media (see, e.g., McKee \& Ostriker 1977; Wolfire \etal\ 1995a).  This
is even the case in the self-gravitating molecular clouds in spirals
given the external pressure term in the virial equation (see Elmegreen
1999).

In the ``standard'' models of thermal phase equilibrium in the neutral
media, the most important heating source is the photoejection of
electrons from the surfaces of interstellar dust grains (Bakes \&
Tielens 1994; Wolfire \etal\ 1995a).  Grains are also important as
potential coolants (e.g., by recombination of electrons and ions on
their surfaces), and in certain conditions may significantly affect
the thermal balance of a medium by the removal of heavy elements from
the gas-phase.  Thus interstellar grains play an important role in the
physics of the ISM in the thin gaseous disks of galaxies.  Their
influence on the thermal state of the gas helps determine the phase
structure of the medium, and grains go on to play an important role in
the star formation process within the dense ISM.

Our understanding of the physical conditions of neutral clouds in the
halos of galaxies is less well developed.  From a theoretical
standpoint, clouds far from the planes of spiral galaxies should
behave similarly to clouds in the disk, with the same heating and
cooling mechanisms and possibly similar phase structure (Wolfire
\etal\ 1995a, b).  Given the importance of dust grains to the thermal
physics of clouds at small heights, \z, from the midplane, knowledge
of the dust content of high-\z\ material is essential if we are also
to understand clouds in the halos and thick disks of spiral galaxies.

Much of the material in the halo and thick disk of a spiral galaxy is
thought to originate in the thin interstellar disk.  The processes
that circulate matter between the thin disks and halos of spiral
galaxies operate on both gas {\em and} dust.  However, depending on
the method of ejection of matter from the thin interstellar disk and
the resiliency of dust grains to destruction by various means (Jones
\etal\ 1994), the dust content of halo material may be different than
that of the disk.  This can have important implications for the
thermal phase structure and physics of high-\z\ gas (Wolfire \etal\
1995a,b).

In our own Galaxy there is evidence from gas-phase abundances that
individual \HI\ clouds at relatively large distances from the plane
contain dust (Sembach \& Savage 1996; Savage \& Sembach 1996, and
references therein).  There is also evidence for the existence of dust
grains in the thickened warm ionized medium or ``Reynolds Layer'' of
the Galaxy (Howk \& Savage 1999a; Lagache \etal\ 1999, 2000).  It
would therefore seem that the processes responsible for lifting the
observed gas clouds out of the plane of the Galaxy also affect the
dust.  And though the dust in the Galactic halo may be to some extent
processed, or partially destroyed, the evidence suggests this
destruction is not total.

Observations of external galaxies have revealed extensive thickened
layers of ionized gas traced by its \Ha\ emission in several edge-on
spiral systems (e.g., Rand, Kulkarni, \& Hester 1990, 1992; Dettmar
1990; Rand 1996; Ferguson, Wyse, \& Gallagher 1996; Hoopes, Walterbos,
\& Rand 1999).  These extraplanar layers of so-called diffuse ionized
gas (DIG) are thought to be similar to the Reynolds Layer of warm
ionized material in the Galaxy (Reynolds 1991).  Among the most
spectacular examples of extraplanar DIG is seen in the nearby
($D\sim9.5$ Mpc) edge-on Sb galaxy \ngc 891 (Dettmar 1990; Rand \etal\
1990; Keppel \etal\ 1991; Rand 1997, 1998).  The observed \Ha\
emission from this galaxy implies an ionized gas layer with multiple
electron scale-heights of $\sim 1.0$ and $\sim2-3$ kpc (Hoopes \etal\
1999; Rand 1997).\footnote{As we will see below, this galaxy contains
an extended dust distribution, with extinction visible to large
distances from the midplane.  Neither of these studies have accounted
for this opacity, which is particularly important for heights $z \la
1.0$ kpc.  The dust effects the observed quantity and distribution of
emission from gas and stars in the thick disk of \ngc 891.}

In Howk \& Savage (1997, hereafter \pI) we intiated a program to study
extraplanar dust in edge-on spiral galaxies.  We presented
high-resolution optical images of the nearby galaxy \ngc 891 taken
with the WIYN 3.5-m telescope at Kitt Peak National Observatory.
These images reveal extensive amounts of highly-structured dust seen
in absorption against the background stellar light of the galaxy.  The
high-resolution WIYN images show hundreds of individual dust-bearing
clouds observable to heights $0.4 \la z \la 1.5$ kpc from the midplane
along the entire observed length of the galaxy.\footnote{Throughout
this work we will use the notation $z$ to denote the distance from the
midplane.  Unless the value in question refers to a single structure,
we will be implicitly assuming $z = |z|$ in our discussion, meaning
that the value $z$ refers to distances from the midplane to either
side of the galaxy.}  A very simple analysis of the radiative transfer
suggests these dust structures are relatively opaque ($A_{V} \ga 0.8$
to 2.0).  Assuming gas-to-dust relationships appropriate for the disk
of the Milky Way (Bohlin, Savage, \& Drake 1978) are applicable to
these clouds, the inferred hydrogen column densities were in excess of
$N_{\rm H} \ga 10^{21}$ atoms \column, implying quite large masses ($M
\ga 10^5$ to 10$^6$ \msun).  Estimates for the potential energies of 
the observed dusty clouds relative to the midplane are in the range
$\Omega \sim \e{52} - \e{53}$ ergs, similar to the energy estimates
derived for Galactic supershells (Heiles 1979).  The implied total gas
mass associated with the ensemble of dusty high-\z\ clouds is roughly
similar to that estimated for the mass of the extraplanar DIG in
NGC~891 ($\sim\e{8}$ M$_\odot$; Dettmar 1990).  The images presented
in \pI\ clearly show a substantial amount of dust is present in the
thick disk of NGC~891.  Subsequent observations by Alton \etal\ (1998)
using the SCUBA submillimeter bolometer have detected $\lambda 850 \
\mu$m emission from extraplanar dust to heights $z \lesssim2$ kpc in
\ngc 891.  The mere presence of the dust features discussed in
\pI\ imply that the mechanism(s) responsible for transporting material
from the thin interstellar disk into the more extended thick disk of
\ngc 891 does not destroy the grains.  We refer the reader to \pI\ for
a detailed discussion of the various mechanisms that may be at work in
lifting gas and dust out of the thin disks of spirals.

In our second paper of this series (Howk \& Savage 1999b, hereafter
\pII), we studied the frequency of such extraplanar dust features in
normal spiral galaxies.  Based on a small WIYN imaging survey of all
the massive, edge-on spirals within $D\la25$ Mpc observable from the
north, \pII\ showed that the presence of extraplanar dust is
relatively common in normal spiral galaxies, with more than half of
the galaxies in our final sample exhibiting prominent extraplanar dust
structures.  Furthermore, in the galaxies of this survey that had
previously been searched for extraplanar DIG, \pII\ showed a
one-to-one correlation between the presence (or absence) of high-\z\
dust structures and the presence (or absence) of high-\z\ DIG.  Thus
galaxies showing extended regions of \Ha\ emission far from the plane
also showed highly-structured dust clouds visible in our broadband
images.

The interpretation of the statistical correlation between the high-\z\
dust structures and the DIG is not clear.  In general those galaxies
showing high-\z\ dust seemed to be relatively uniform in the
properties of the dust-bearing clouds, while the \Ha\ morphologies and
brightnesses showed significant variation (see Rand 1996 and Pildis,
Schombert, \& Bregman 1994).  The discussion presented in
\pII\ argued that the extraplanar dust features seen in these edge-on
spirals likely trace a dense phase of the ISM at high-\z, separate and
physically distinct from the extraplanar DIG.  Several lines of
evidence were used to argue for such a configuration, though this
conclusion was primarily based on the rough column density estimates
for the dusty structures and on the general lack of morphological
similarity between the dust-bearing and \Ha -emitting material.  The
implied column densities with the observed dimensions of the
structures yielded crude estimates of the particle density that were
irreconcilable with the expected densities of the extraplanar DIG.
The DIG at $z \sim 1$ kpc is expected to have local electron densities
of order $n_e \sim 0.2-0.3$ assuming a volume filling factor of
$f\sim0.2 - 0.25$ (Rand 1997).  The dust features are only detectable
in these images because they are significantly denser than their
surroundings.  Estimates of the particle densities in the extraplanar
clouds traced by the dust absorption in the galaxies observed in \pII\
are in the range $\sim1 - 10$ \percc.

While there is now evidence for a statistical relationship between the
the presence of extraplanar \Ha\ emission and high-\z\ dust
absorption, there is as of yet little information on any physical
relationship between the DIG and dust within an individual galaxy.
Are these observational tracers measuring the same high-\z\ material,
or do they each give insight into a separate phase of the thick disk
or halo ISM?  In this paper we present much deeper observations of
\ngc 891 in the BVI bands than the data of \pI.  We also present deep,
high-resolution narrowband (\Ha +[\ion{N}{2}]) images of ionized gas
emission from this galaxy.  The purposes of these observations are to
probe the known dust structures in this galaxy as far from the
midplane as possible, and to directly investigate the physical
relationship between the dust features and the DIG in the disk-halo
interface of
\ngc 891.


Our presentation is arranged as follows.  In \S
\ref{sec:observations} we present a summary of our observations
and data processing.  This includes a description of our unsharp
masking technique.  The results of our broadband imaging, and the
implications for the extraplanar dust structures, are presented in \S
\ref{sec:thickdisk}.  Our \Ha\ images are discussed in \S
\ref{sec:halpha}, including a direct comparison of the \Ha\
emission from \ngc 891 and the morphology of the absorbing dust
structures.  The implications of the observations, and in particular
the comparison of the \Ha\ and dust morphologies, are discussed in \S
\ref{sec:discussion}.  That comparison shows that these two observational
tracers, the dust absorption and the \Ha\ emission, are revealing two
different interstellar phases in the multiphase halo of \ngc 891.  We
summarize the major results of this work in \S \ref{sec:summary}.

\section{Observations and Reductions}
\label{sec:observations}

\subsection{Observations and Processing}

All of the observations presented here were obtained with the WIYN
3.5-m telescope at Kitt Peak National Observatory in the fall of 1997,
though we have also included some of the data from \pI\ in our
analysis, which were taken 1996 December 4 (UT).  A log of our
observations is given in Table \ref{table:log}.  This table shows
the filter used, the date of observation, and the seeing-limited
resolution, expressed as the FWHM (in arcseconds) of Gaussian fits to
the stellar images using the IRAF\footnote{IRAF is
distributed by the National Optical Astronomy Observatories, which is
operated by the Association for Research in Astronomy, Inc., under
cooperative agreement with the National Science Foundation.} routine
{\tt IMEXAM}, for each exposure.  The images presented here were taken
under non-photometric conditions.  Our analysis depends only on
measures of the relative intensity and does not require photometric
calibration (see \pI ).

The WIYN imager in use at the time of our observations is a thinned
2048$\times$2048 STIS CCD with 21 $\mu$m pixels.  Placed at the f/6.5
Nasmyth focus of the WIYN telescope, the imager has a $6\farcm7 \times
6\farcm7$ field of view, with each pixel subtending $0\farcs196$ on
the sky.  The images have been bias-subtracted and flat
field-corrected in the usual manner within IRAF.  

The flat fields were derived from observations of the ``Great White
Spot'' in the telescope dome.  In principle subtle variations in the
flat-field of the image could mimic dust absorption in NGC~891 (though
very few of the dust features we observe can be referred to as
``subtle'').  Our experience suggests that the dome flats can
adequately remove small-scale structures associated with the CCD.  In
the course of our observing we moved the center of the CCD relative to
the galaxy by $2\arcsec - 30\arcsec$ with each new exposure.  This
shifts any instrumental flat field features relative to the galaxy.
Each individual image was registered to a common reference frame using
the positions of foreground stars.  The rms errors in the alignment of
the images are typically $\la5\%$ of a pixel.  In the resulting
coadded data, the effects of any artifacts associated with the flat
field structure of the CCD should be greatly diminished.  We have
found no real evidence, when comparing our final coadded images with
the individual exposures, for confusion of flat-field features with
absorbing dust structures.

Table \ref{table:finalimages} gives the properties of our final
coadded images, including the final angular resolution of the images.
Each image was smoothed with a Gaussian to the worst resolution in a
given band before coaddition.  For comparison, at the distance of \ngc
891 ($\sim9.5$ Mpc; Tully 1988), $0\farcs8$ corresponds to $\sim37$
pc.  The narrowband images will be described more fully in \S
\ref{subsec:narrow}.  Removal of cosmic rays in the images was
performed using standard sigma-clipping algorithms within IRAF.  We
have derived the astrometric plate solution for our images using a
grid of stars whose coordinates were measured in the Digitized Sky
Survey.  The rms error in using the plate solution to derive
coordinates is approximately 1\farcs0.  The sky background in each
frame is estimated by fitting a planar (or in a few cases quadratic)
surface to regions of the images free from light of the galaxy.  Stars
in these regions have been removed using a median smoothing technique.
This approach is somewhat problematic given the galaxy fills such a
large fraction of the CCD area.  The sky-subtraction is not perfect,
but we have endeavored to make as reliable a fit as possible.


\subsection{Image Display and Unsharp Mask Procedure}
\label{subsec:unsharpmask}

All of the results presented in this work require an effective yet
accurate method of displaying our high-quality images.  We will rely
very heavily on unsharp masked displays of our data to show the
multitude of structures present at high-\z\ in \ngc 891.  Our approach
to producing quality unsharp masked displays is discussed in some
detail in \pII, but we present here the principal points of that
discussion.

\begin{figure}
\epsscale{1.0}
\plotone{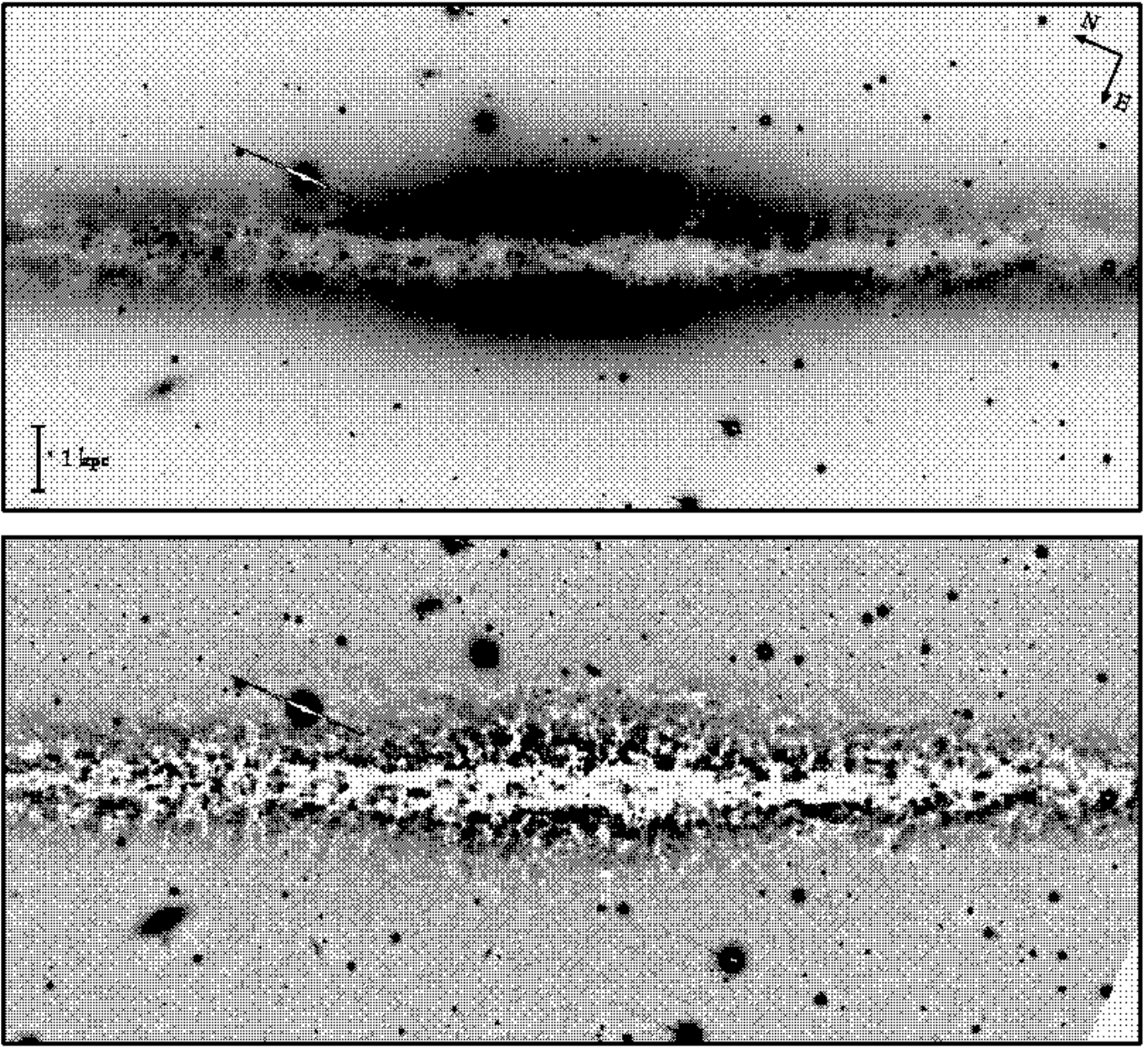}
\caption{WIYN V-band images of \ngc 891.  The top panel shows the
V-band image, while the bottom panel shows the unsharp masked version
of this image.  The display is inverted such that darker regions
represent brighter emission.  Regions of dust extinction are lighter
than their surroundings.  This image covers $6\farcm4 \times 2\farcm8$
($17.3 \ {\rm kpc} \, \times7.6$ kpc); a scale bar denoting 1 kpc is
shown.  North and east are marked.  Note that this display is flipped
across the y-axis from the figures in \pI\ (e.g., Figure 2 of \pI) to
be consistent with most other published images.  The dust absorbing
structures provide a direct view of the dense neutral medium in the
multiphase halo of \ngc 891.
\label{fig:Vfull}}
\end{figure}

Figure \ref{fig:Vfull} shows our final V-band data for NGC~891, as
well as the unsharp masked version of the data.  The V-band image
(top) is displayed to show the distribution of stellar light along the
length of the galaxy.  With this display the light from the bulge is
saturated, making it difficult to identify high-\z\ dust structures.
However, the unsharp mask of the V-band data (bottom) allows us to
show the absorbing structures along the entire length of the galaxy in
one display.  This masked version of the image shows the complexity of
the absorbing structures that thread through the bulge area.

We derive the unsharp masked versions of our images by dividing the
original image by a version smoothed with a Gaussian kernel having a
FWHM of 35 pixels ($6\farcs9$).  With this approach we remove the
large-scale gradients in the background light of the galaxy, allowing
us to display all parts of the galaxy in a uniform manner.  The
procedure tends to accentuate structures on scales smaller than
smoothing kernel ($\Delta \la 310$ pc at the distance of \ngc 891).
This can be both an advantage and a disadvantage.  Most of the
structures we find have minor axis scales similar or smaller than
this, though not all.  In some sense this offsets the natural tendency
to have one's eye drawn towards very large structures.  Although we
tend to identify structures only in computer displays of the science
images (i.e., not the unsharp masks), the reader's eye may be drawn
more to the small-scale structures in our unsharp masks.  As in all of
our papers, the unsharp masked images are only used for display, never
for quantitative measurements.

Bright stars near \ngc 891 can produce large artifacts in unsharp
masked images, destroying information over a disproportionately large
area.  We have produced our unsharp masks by dividing the original
data by a smoothed image with such problem stars replaced by a two
dimensional fit to the surrounding background light before the
smoothing process. The replaced area is usually chosen to be a
circular aperture of several times the FWHM of the seeing disk.  For
fainter examples, the aperture radius was typically twice the size of
the seeing disk, while the brighter stars often required apertures in
excess of five times the FWHM of the seeing disk.  We have also
removed the effects of CCD blooming near a few bright stars.

By dividing the original image by the smoothed image with bright stars
removed, we are able to keep such stars from influencing large regions
of the galaxy, while at the same time showing where these stars lay
with respect to the target.  Artifacts can be present in our unsharp
masked images very near the positions of the brightest stars.  Faint
halos (white in this display) can be seen around a few of the
background galaxies and faint stars in Figure \ref{fig:Vfull}.

While there are potential problems associated with the interpretation
of the unsharp masks displayed here, this approach is much more
effective at showing the reader the true extent and complexity of the
dust features we see at high-\z\ in \ngc 891.  We are confident that,
with the exception of some structures visible very near the brightest
stars, the dust features seen in our unsharp masked images are real
and not artifacts of the masking procedure.

\subsection{Narrowband Emission Line Images}
\label{subsec:narrow}

We will present narrowband images of \ngc 891 to study the
distribution of ionized gas in this galaxy.  To derive an image of the
emission lines (\Ha + [\ion{N}{2}]) from this galaxy, we have taken
images through a narrowband filter centered on the \Ha\ emission line,
and filters that cover continuum regions on each side of the \Ha\
wavelength region.  The characteristics of the on- and off-band
filters used are given in Table \ref{table:narrowfilters}.  The filter
Off 1 is from the Kitt Peak filter set (KP 809), while the \Ha\
on-band filter and the filter designated Off 2 are from the WIYN
filter library (W015 and W016, respectively).

The \Ha\ on-band filter contains emission from the \Ha\ and nearby
[\ion{N}{2}]\footnote{Our emission line image contains
contributions from both \Ha\ and the nearby forbidden [N$\,$II] lines.
For brevity's sake we will hereafter refer to our image as an \Ha\
image, though the reader should be aware that the bandpass also
contains the [N$\,$II] transitions.} lines plus a stellar continuum
contribution.  To produce a pure emission line image we must subract
off the contribution of the stellar continuum.  Our two off-band
filters cover wavelength ranges shortward and longward of the \Ha\
emission line.  To derive an appropriate continuum image we have
coadded the off-band data with an appropriate weighting factor.  Each
of our off-band images was normalized to one second exposure time and
combined with a weighting factor, W, defined as
\begin{equation}
W \equiv [ T \, \times \, |\lambda_{cen} - \lambda_{\rm o}| ]^{-1},
\end{equation}
where $T$ is the average transmission of the filter, $\lambda_{cen}$
is the transmission-weighted average wavelength of the filter, and
$\lambda_{\rm o}$ is the rest-wavelength of the \Ha\ emission line.
The ratio of the weighting factors for the two off-band filters is
$W({\rm Off \ 1}) / W({\rm Off \ 2}) = 0.81$.


After combining the two off-band filters to produce a single continuum
image, we have performed aperture photometry of a grid of $\sim20$
stars in the sky-subtracted on- and off-band images.  The fluxes
derived from these stars were used to determine the relative scaling
between the on-band and continuum images.  The continuum image was
multiplied by this scale factor and subtracted from the on-band image
to produce the final emission line image.  The use of this scale
factor assures that the foreground stars in our image (which should be
pure continuum sources) are completely subtracted.  Stars which are
saturated, however, do not subtract completely.

We reiterate that the size of the galaxy relative to the WIYN imager
is large, and our sky subtraction is not perfect.  In particular the 5
kpc layer detected by Rand (1997, 1998) cannot be reliably
characterized in our images given its extent relative to the size of
the CCD.  We do detect emission to $z \sim4$ kpc in our \Ha\ image,
but the true amount and distribution of such high-\z\ gas is confused
by the background subtraction uncertainties.

\subsection{Dust Feature Nomenclature and Measurements}
\label{subsec:nomenclature}

\begin{figure}
\epsscale{0.75}
\plotone{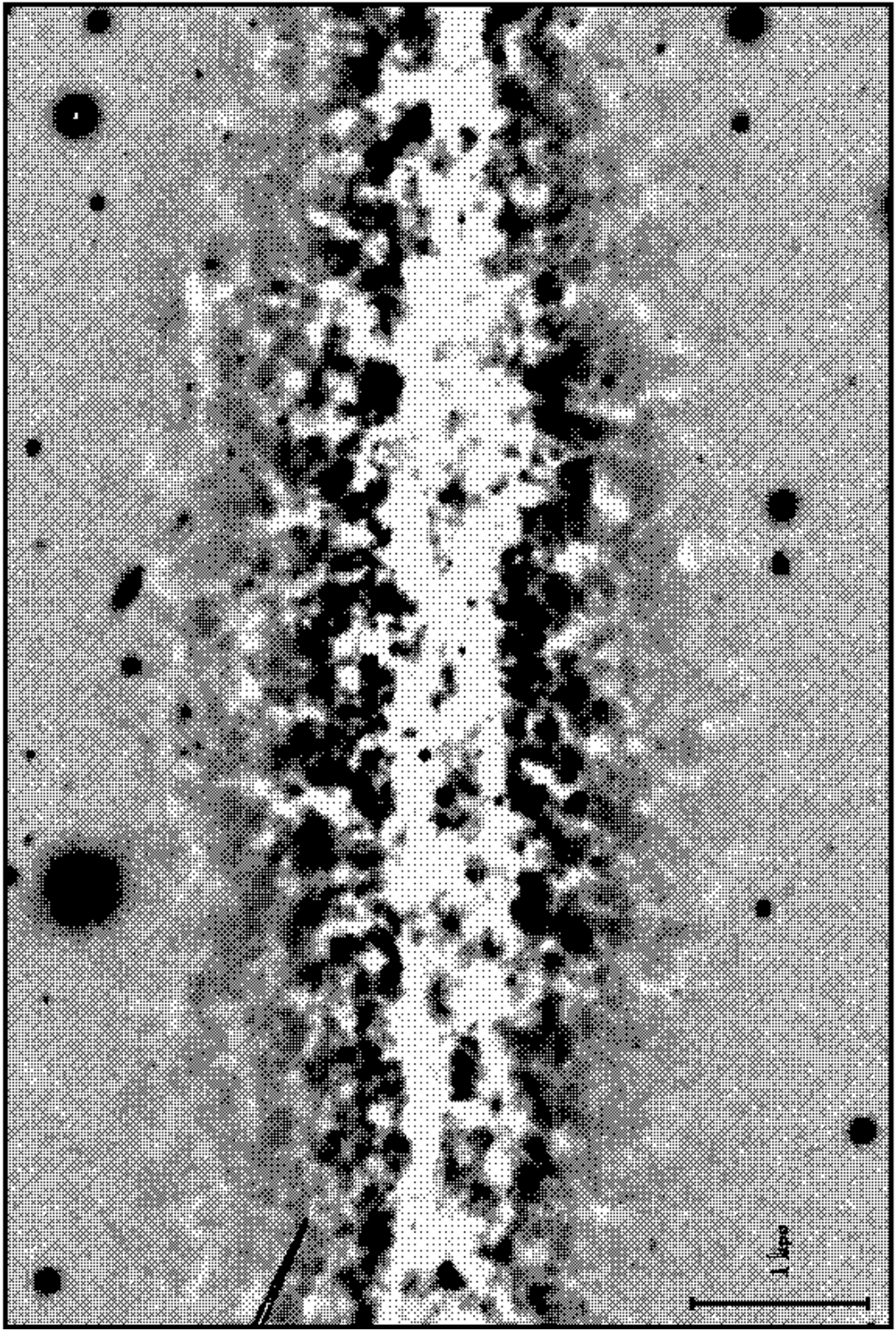}
\caption{An expanded view of the unsharp masked V-band data for a
section of the central $2\farcm7 \times 1\farcm8$ (or $7.3 \ {\rm kpc}
\, \times4.9$ kpc) of \ngc 891.  A 1 kpc scale bar is displayed in the
lower left corner of the image.  
\label{fig:Vmid}}
\end{figure}

\begin{figure}
\epsscale{0.75}
\plotone{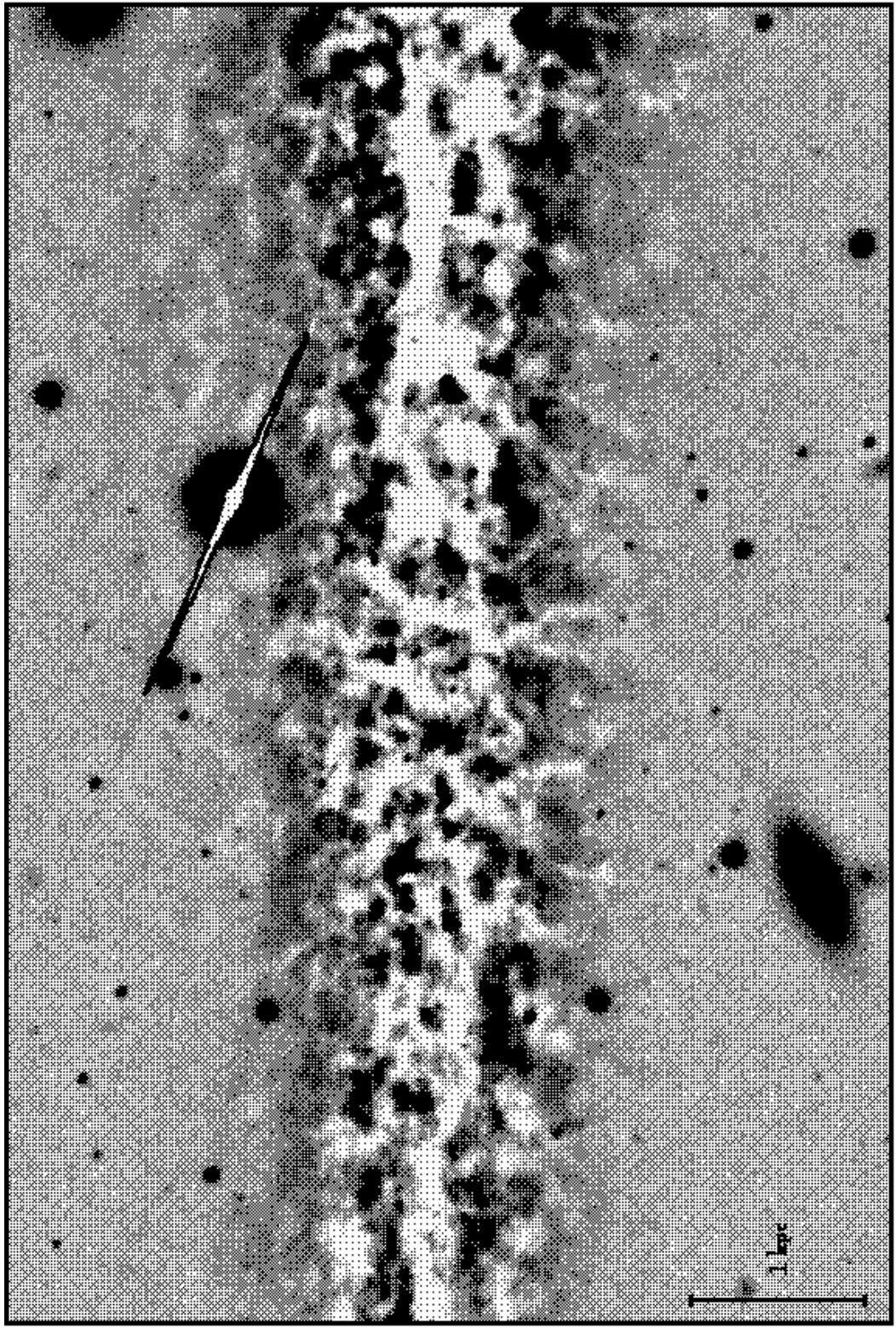}
\caption{As Figure \ref{fig:Vmid} but showing a section of the disk of
\ngc 891 centered to the northeast of the nucleus.
\label{fig:Vne}}
\end{figure}

\begin{figure}
\epsscale{0.75}
\plotone{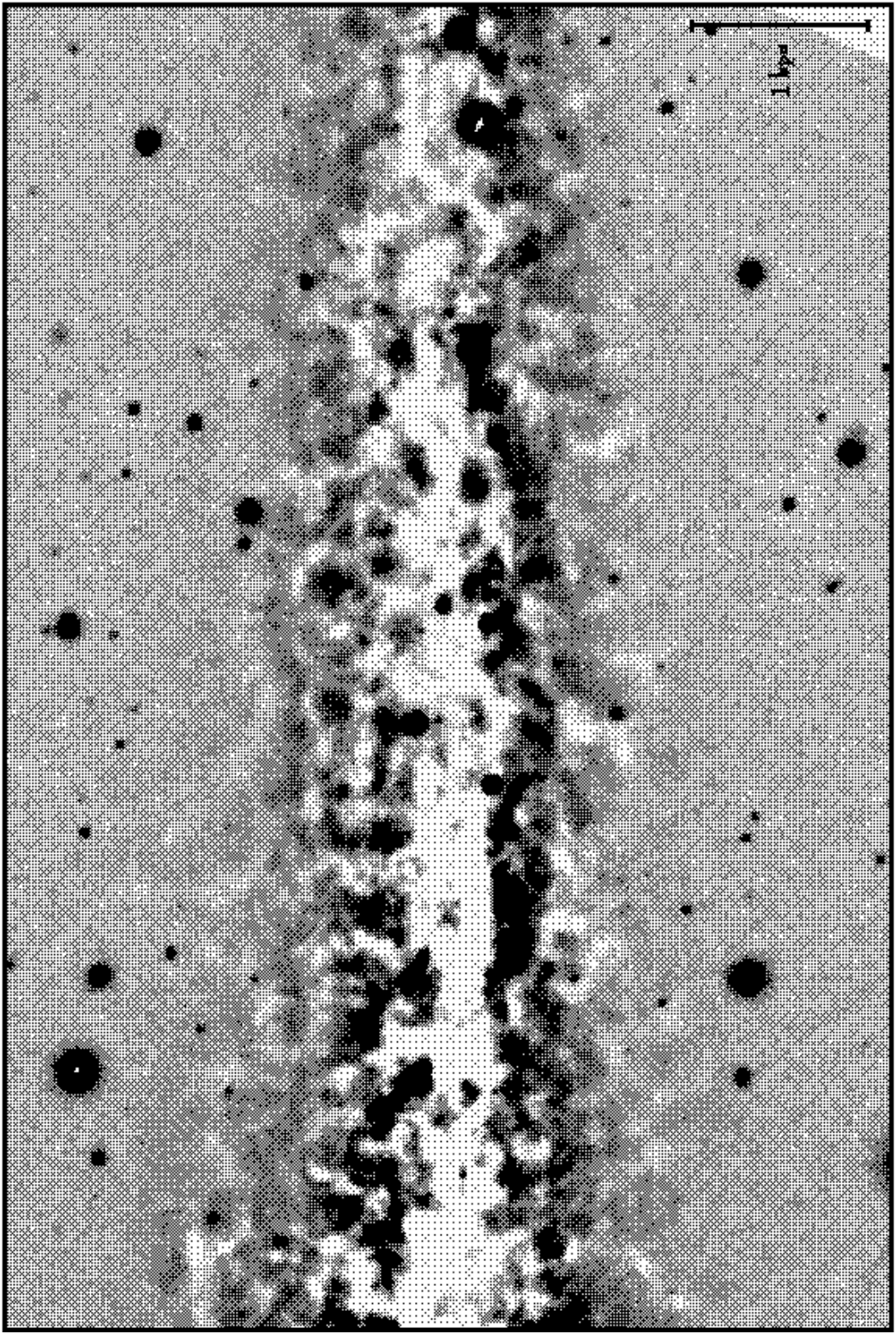}
\caption{As Figure \ref{fig:Vmid} but showing a
portion of the disk to the southwest of the nucleus.
\label{fig:Vsw}}
\end{figure}

\begin{figure}
\epsscale{0.75}
\plotone{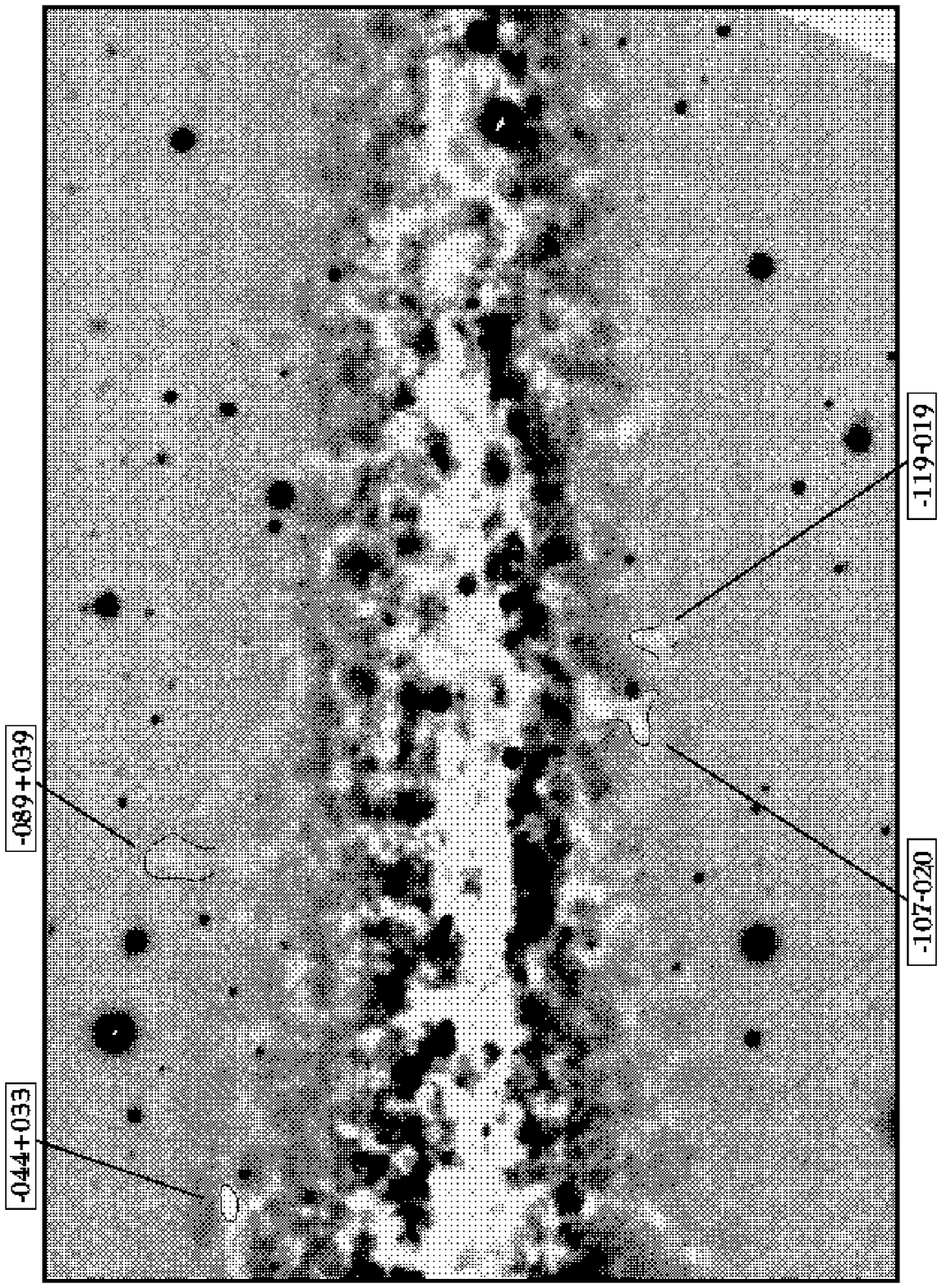}
\caption{The same region as Figure
\ref{fig:Vsw}, but with several individual absorbing structures
identified.  The properties of these clouds are summarized in Tables
\ref{table:observed} and \ref{table:derived}.
\label{fig:IDsw}}
\end{figure}

Figures \ref{fig:Vmid} through \ref{fig:IDsw} show close-up
views of the disk in three sections.  Figures \ref{fig:Vmid},
\ref{fig:Vne}, and \ref{fig:Vsw} show the unsharp masked views
of the central, northeast, and southwest sections of \ngc 891,
respectively.  Each figure shows a $2\farcm7 \times 1\farcm8$ or $7.3
\ {\rm kpc} \, \times 4.9$ kpc section of the disk.  Figures 
\idfigs\ show these same sections with several individual dust 
features identified.  Following \pII\ we assign each structure an
identification of the form NGC 0891:D $\pm$XXX$\pm$ZZZ.  The ``D''
denotes a dust cloud.  The values XXX and ZZZ are the projected
distances (in arcsec) of the features from the center of the galaxy
traced along the major and minor axes, respectively.  North and east
are positive in this reference frame.  We assume the center of \ngc
891 to be $\alpha_{2000} = 2^h \, 22^m \, 33\fs3$, $\delta_{2000} =
+42^\circ \, 20\arcmin \, 52\farcs5$ (a mean of the de Vaucouleurs
\etal\ 1991 and NED\footnote{The NASA/IPAC Extragalactic Database
(NED) is operated by the Jet Propulsion Laboratory, California
Institute of Technology, under contract with the National Aeronautics
and Space Administration.}  coordinates\footnote{We found neither
individual measurement was in agreement with our images; though the
mean matched the optical center of the galaxy in our images quite
well.  }) and a position angle for the disk ${\rm PA} = 23\fdg5$ east
of north (\garciaburillo\ \etal\ 1992).  The designations are only
meant to be approximate, and the equatorial coordinates should be used
for making detailed comparisons with other wavebands.  Hereafter we
will typically abbreviate these identifications $\pm$XXX$\pm$ZZZ.

Table \ref{table:observed} contains the equatorial coordinates,
$z$-heights, and approximate dimensions of each structure identified
in Figures \idfigs.  We have roughly outlined each of the identified
structures to help the reader identify the regions used in our
calculations.  In a few cases we mark with dashed lines regions of
absorption that may be associated with the identified structure but
are not used in our measurements of the structure.

In \pI\ we characterized dust structures at high \z\ in NGC~891 by
their ``apparent extinction,'' $a_\lambda$.  We defined the apparent
extinction in a waveband $\lambda$ as
\begin{equation}
 a_\lambda = -2.5 \log( S_{dc, \lambda} / S_{bg, \lambda} ),
\label{eqn:extinct}
\end{equation}
where $S_{dc, \lambda}$ is the surface brightness measured towards a
dust cloud, and $S_{bg, \lambda}$ is the surface brightness of the
local background (see \pI).  The light measured towards a given dust
feature contains an extincted component of starlight originating
behind the feature along the line of sight as well as an unextincted
component emitted in front of the feature.  The apparent extinction in
a waveband, $a_\lambda$, is a lower limit to the true extinction,
$A_\lambda$, given the unextincted star light and the unknown
contribution of scattering into the beam (see discussion in \S
\ref{subsec:properties}).  Table \ref{table:observed} contains 
average values of $a_B$, $a_V$, and $a_I$ for each feature (in
magnitudes), measured as described in \pI.

The combined statistical uncertainties and spatial variations in the
dust structures give dispersions typically of order $0.05 - 0.10$ mag
for the average $a_\lambda$ values given in Table
\ref{table:observed}.  The systematic errors in the $a_B$ and $a_V$
determinations could be larger.  In the B and V images there is
significant overlap between the individual dust structures. This
overlap often makes the definition of a true continuum or background
difficult: at low \z\ the dust so completely covers the galaxy that
the definition of a continuum or background for making these
measurements is uncertain.  Measurements for clouds much below $z
\sim1.0$ kpc can be difficult to make, and the errors on the shorter
wavelength apparent extinctions can be great.  At larger \z, where the
dust features become less common and typically show lower $a_\lambda$
values, the systematic uncertainties are likely less important than
the measurement error and spatial variations within an individual
cloud.  The overlap of the multitude of dust features is less
problematic in the I-band, making the $a_I$ measurements more secure
than the shorter wavelength apparent extinction values.

\section{The Dusty Thick Disk of NGC~891}
\label{sec:thickdisk}

Our new broadband images reveal a wealth of absorbing structures
stretching to very large distances from the midplane of \ngc 891.
Figure \ref{fig:Vfull} shows our V-band image of \ngc 891 and the
unsharp masked view of these data.  Though we only show the V-band
images here, the structure of the dust in the other two wavebands is
much the same as that seen in the V-band.  The absorbing structures
are more prominent in the B-band and less so in the I-band.  The
V-band images represent a good compromise between the larger opacity
at shorter wavelengths and the generally higher signal to noise in the
longer wavelength observations.

\begin{figure}
\epsscale{0.75}
\plotone{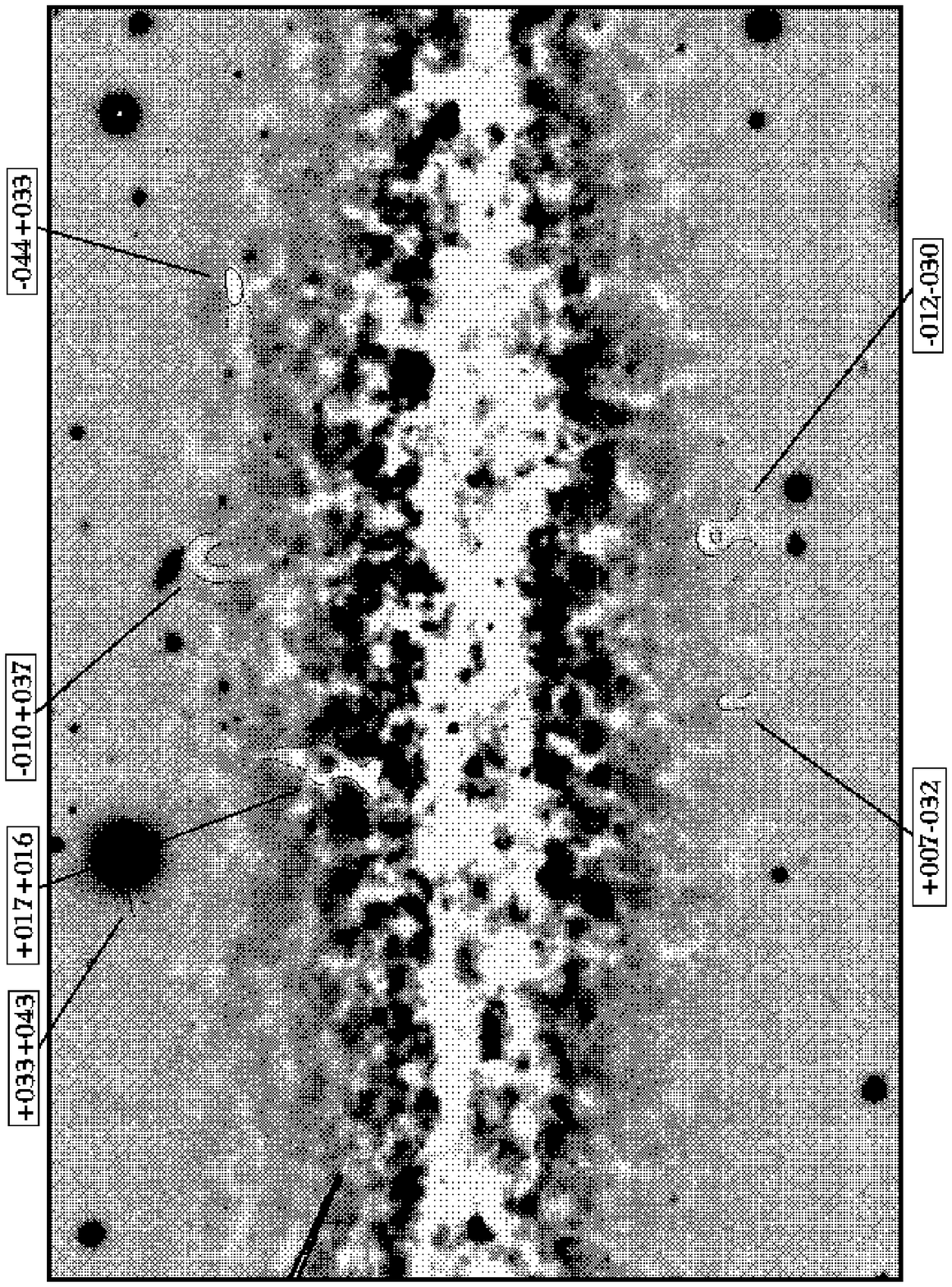}
\caption{The same region as Figure \ref{fig:Vmid}, but with several
individual absorbing structures identified.  The properties of these
clouds are summarized in Tables \ref{table:observed} and
\ref{table:derived}.
\label{fig:IDmid}}
\end{figure}

\begin{figure}
\epsscale{0.9}
\plotone{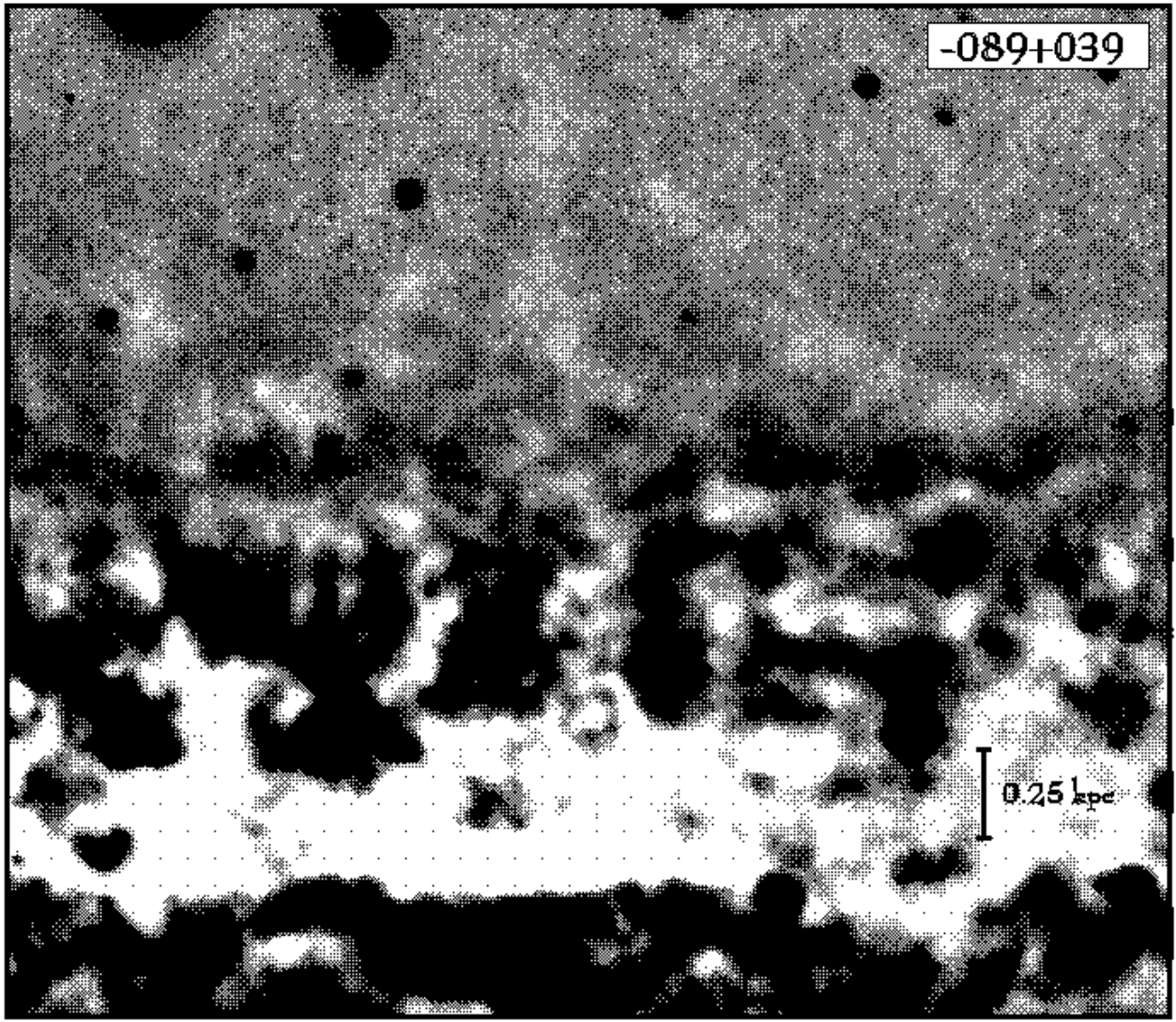}
\caption{ A close-up view of the unsharp masked V-band data showing
NGC 0891:D $-089+039$.  This structure as identified in Figure
\ref{fig:IDsw} is only the top-most cloud centered at $z \sim 1850$
pc.  However, this cloud may be part of a much larger structure
traceable almost to the midplane.  Absorption from material aligned
with this structure can be traced to $z\sim2.0$ kpc, and possibly
higher.  A scale bar of 0.25 kpc length is shown.
\label{fig:filament1}}
\end{figure}

The primary difference between the current dataset and that presented
in \pI\ is the depth of the observations.  The new broadband data are
a factor of $\sim15$, on average, deeper than the data presented in
\pI, though the resolution of the current dataset is slightly worse.  
This allows us to identify structures against the background stellar
light to larger heights from the midplane.  We find firm evidence for
structures at $z \sim2.0$ kpc along most of the length of the galaxy.
For example, Figure \ref{fig:filament1} shows a close-up view of the
structure identified as $-089+039$ in Figure
\ref{fig:IDsw}, which will be discussed in more detail below.
Material that may be associated with this structure can be traced to
at least $z\sim2.0$ kpc, if not higher.  Another discrete feature at
quite large \z\ is identified as $+033+043$ in Figure \ref{fig:IDmid}.
This structure is seen prominently against the light of a background
elliptical galaxy (approximately E0, centered at $\alpha_{2000} = 2^h
\, 22^m \, 30\fs5$, $\delta_{2000} = +42^\circ \, 21\arcmin \,
35\farcs7$) at a height $z\sim2.0$ kpc.  There are a few individual
structures that can be identified when examining the one dimensional
light distribution at heights as large as $z\sim2.0-2.5$ kpc.

The dust structures are much less prevalent at heights beyond $z \ga
1.8 - 2.0$ kpc than at lower \z.  Though the background stellar light
is rapidly dimming with height, it is clear there are no structures at
heights in excess of 2.0 kpc like those seen at heights $z \sim
0.5-1.5$ kpc.  Our images show the clear signature of the stellar
thick disk (see Morrison \etal\ 1997 and Morrison 1999) in all three
wavebands.  To confirm the apparent thick disk light in our images was
not simply an artifact of the extended wings of the telescope point
spread function (PSF), we have fit several bright stars with a Moffat
function (Moffat 1969) to approximate the instrumental PSF.  We have
then constructed several realistic thin disk model light distributions
and convolved these with the instrumental PSF.  We find at most a few
percent of the light observed to heights $z > 3-4$ kpc could be the
result of scattering of light from the thin disk \ngc 891.  We have
found similar results when including a lower scale-height thick disk
than that fit by Morrison \etal\ (1991), e.g., by adopting scale
heights $\la 1$ kpc.  Thus there is detected background light against
which we could view the dust features to large \z -heights (e.g.,
$z\ga 3$ kpc in the V-band).  While a few features can be identified
in the range $z\sim2.0-2.5$ kpc, the number of structures at these
heights is much less than at lower heights, and the apparent
extinctions are very small (typically $a_V \la 0.1 - 0.15$; often
unmeasurable in the other two bands).


Color maps of \ngc 891 derived from our BVI images give a similar
picture: we find \ngc 891 gets {\em bluer} with increasing height
until $z\sim1.9$ kpc or so.  The effects of individual dust clouds on
the vertical light distribution become significant in one-dimensional
cuts through our data at heights in the range $z\sim 1.8 - 2.0$ kpc.
%
This is consistent with a phase change of the material, where dust at
heights in excess of this value is more diffusely distributed and
hence not visible in our images (see \S \ref{sec:discussion}).
Although, the lack of dust structures at heights beyond $\sim 1.8 -
2.0$ kpc may simply be tracing the maximum extent to which the
material was ejected.

We believe the paucity of structures at very high \z\ ($\ga 2.0$ kpc)
is a real effect.  There is not a large amount of highly-structured
(clumped) dust visible in our images at these heights; however, we
cannot rule out the existence of a more smoothly distributed component
of the dust at very high \z.  Our images reveal only the dusty
structures more opaque than their surroundings.  The visible
structures are likely tracing significantly denser regions than the
surrounding material.  The decreasing number counts and apparent
extinctions of features at very high \z\ do not imply there is not
dust at these heights, but it is clear there are very few of the
overdense, dusty clouds seen at low \z\ to be found at heights in
excess of $z\ga2.0$ kpc.  


In \pI\ we found the number of dust features on either side of the
midplane of \ngc 891 was roughly similar.  This suggests the dust
structures do not arise in a warp along the line of sight, which would
cause an asymmetry in the distribution of absorbing structures.  The
present images show that though the number of structures is similar
for each side of the midplane, there is an interesting difference in
the morphology of the highest-\z\ absorbing structures from one side
of the galaxy to the other.  The northern side of the galaxy (up in
our images) seems to show an extended series of structures oriented
roughly parallel to the plane of the galaxy at $z\sim1.5 - 1.8$ kpc.
The structure $-044+033$ in Figure \ref{fig:IDmid} is indicative of
the dominant orientation of features at these heights.  To the south
of the midplane, however, the structures are oriented primarily
vertically near $z\sim1.3 - 1.7$ kpc.  These vertically-oriented
structures show an open geometry to high \z, with a few cometary-like
structures (e.g., $-012-030$ and $+007-032$ in Figure
\ref{fig:IDmid}).  The open structures on the southern side of \ngc 891 
are possibly tracing galactic outflow or inflow, though other causes
may be viable as well.

It is possible that the ``flattening'' of structures on the northern
side of the galaxy is caused by the ram pressure from the motion of
\ngc 891 through an intragroup medium.  The \ngc 1023 group to which
\ngc 891 belongs is dominated, in mass and luminosity, by three disk
galaxies (Tully 1980): \ngc 1023 (SB0), \ngc 891 (Sb/SBb?), and \ngc
925 (SBc).  The total extent of the group is $\sim1.6 \ {\rm Mpc}
\, \times3.3$ Mpc.  \ngc 891 is 0.78 Mpc projected distance from \ngc
1023, similar to the separation between the Milky Way and Andromeda,
and 1.5 Mpc from \ngc 925.  The group as a whole is likely bound
(Tully 1980; Hart, Davies, \& Johnson 1980), and \ngc 891 lies near
its periphery.  It is not unreasonable to expect that a group composed
of three moderately massive disk galaxies with an associated
population of dwarf galaxies (at least 10; see Tully 1980) should
contain an intragroup medium.  The communication of \ngc 891's
movement through a hypothetical medium could propagate through the
higher-\z\ material to form the horizontally-oriented structures on
the northern side of the galaxy.  

Other causes for the flattening of the dust structures on the northern
side of \ngc 891 may be viable as well, such as a slight warp of the
disk along the line of sight.  However, Swaters, Sancisi, \& van der
Hulst (1997) have ruled out a prominent line of sight warp in \ngc 891
by comparing models with their \HI\ channel maps.  Though the
morphological differences between the two sides of this galaxy may be
interpreted in many ways, the possibility that the flattening of the
high-\z\ structures on the northern side is an intriguing possibility.

\subsection{Physical Properties of Individual Dust Structures}
\label{subsec:properties}

The directly observable physical properties of a subset of dust
features are given in Table \ref{table:observed}.  From these
observables we can infer rough information on the column density,
mass, and gravitational potential energies of the individual
structures.  In \pI\ the radiation transfer through an individual dust
structure was treated very simply.  We assumed some fraction $x$ of
the stellar light of the galaxy resided between the observer and the
dust structure.  Therefore, a fraction $(1-x)$ of the background
stellar light was extincted by the dust feature.  In this case the
ratio of the observed surface brightness towards a dust cloud, $S_{dc,
\lambda}$, to that of the local background  $S_{bg, \lambda}$, could be 
written
\begin{equation}
S_{dc, \lambda} / S_{bg, \lambda} = x_\lambda + (1-x_\lambda)
\exp[-\tau_\lambda(x_\lambda)],
\label{eqn:radtrans}
\end{equation}
where $\tau_\lambda(x_\lambda)$ is the estimated optical depth in a
waveband $\lambda$.  If $x_\lambda$ is chosen correctly and scattering
is not important, then $\tau_\lambda = \tau_\lambda(x_\lambda)$, where
$\tau_\lambda$ is the ``true'' absorption optical depth.  Using
observations in three wavebands and assuming $x_\lambda$ to be
constant with wavelength ($x_\lambda = x$), we derived rough
corrections to account for the stellar light arising in front of the
dust structures.  We will denote these corrected values $a_\lambda (x)
\equiv 1.086 \tau_\lambda(x)$, where we have derived some 
fraction $x$ of the stellar light to arise in front of the dust
structures.  The corrected apparent extinctions $a_V (x)$ were derived
in \pI\ assuming the ratios of the optical depths in the three
wavebands followed a Cardelli, Clayton, \& Mathis (1989)
parameterization of the interstellar extinction curve.  The
first-order corrected $a_V (x)$ values are lower limits to the true
extinction, $A_V$, given the unknown geometry, and hence the unknown
contribution from scattering.  The corrected $a_V(x)$ values derived
in this way are therefore more correctly called an ``attenuation''
(e.g., Ferrara \etal\ 1999; Meurer, Heckman, \& Calzetti 1999).  If
$x$ is roughly correct, we have a relationship $a_V \equiv a_V(0) <
a_V(x) < A_V$.

Including the observed values of $a_I$ for the current set of dust
features in this approach, however, does not yield a consistent
solution for $x$ and $a_V(x)$ for reasonable values of $R_V$.  We
believe the reasons for this are (at least) two-fold.  The
aforementioned differences between the structure of the features when
viewed in the B- and V-band data versus that seen in the I-band data
imply that the $a_\lambda$ measurements of individual dust features
may be probing different regions of the clouds at different
wavelengths.  Further, the assumptions made in the very simple
approach to the radiation transfer presented in Equation
(\ref{eqn:radtrans}) are likely not valid.  The effects of scattering
may be important.  Previously the scattered light could be
approximately treated by incorporating it into the value of $x$.
However, it is our belief that the large wavelength range covered by
the current set of observations shows the assumed wavelength
independence of $x$ is erroneous, at least as formulated in Equation
(\ref{eqn:radtrans}).

The values of $a_V \equiv a_V(0)$ given in Table \ref{table:observed}
are firm lower limits to the true extinction, $A_V$, through the
individual dust clouds.  In Table \ref{table:derived} we show the
magnitude of the implied correction by giving $a_V (x)$ for values of
$x=0.0,$ 0.25, and 0.5.  The $a_V (0) \equiv a_V$ measurements are
firm lower limits to the true extinction, and we use these in all of
our subsequent calculations.

We can use the lower limits $A_V > a_V (0)$ to roughly estimate
several other physically interesting properties of the individual dust
features.  Assuming the observed relationship between total hydrogen
(neutral plus molecular) column density and color excess derived by
Bohlin \etal\ (1978) for the Galactic disk is appropriate for the
high-\z\ clouds in \ngc 891, we can estimate the hydrogen column
density, $N_{\rm H}$, of the high-\z\ dust features with:
\begin{equation}
N_{\rm H} >  1.9 \times 10^{21} \ a_V(0) \ 
        {\rm [cm^{-2}]}
\label{eqn:columndensity}
\end{equation}
(e.g., \pII).  With an estimate for $N_{\rm H}$ and the projected
surface area of a cloud, we can estimate the enclosed mass in the
observed dust structures.  The results of these calculations are given
in Table \ref{table:derived}.  The outlines of the dust clouds in
Figures \idfigs\ identify the areas used for calculating the mass of
each structure.  We include a factor of 1.37 correction to the mass
estimates for helium and the heavy elements.  Assuming rough values
for the distribution of mass in \ngc 891 we estimate the gravitational
potential energy, $\Omega$, of the structures relative to the
midplane.  We assume the mass is distributed in an isothermal sheet
[mass density $\rho \propto {\rm sech}^2 (z/2 z_o)$] with a mass scale
height $z_o\sim0.35$ kpc (e.g., the near infrared observations of Aoki
\etal\ 1991 which roughly agree with the results of Xilouris \etal\
1998, who fit exponential disk models to the light distribution) and
mass surface density $\rho_o = 0.185$ \msun pc$^{-3}$ (derived for the
solar neighborhood by Bahcall 1984).  The potential energies of the
structures relative to the midplane are then calculated as in \pI,
using the \z -heights and masses from Tables \ref{table:observed} and
\ref{table:derived}. These estimates are order of magnitude values
and should be viewed with some caution.

The values $a_V (0)$ given in Table \ref{table:derived} are in the
range 0.2 to 0.8 mag.  It is unlikely that the true values of the
extinction are anywhere near this small.  However, even adopting the
smallest allowed values of $a_V(x)$, the estimates of $N_{\rm H}$,
mass, and energy are still large.  These values are derived assuming
the gas-to-dust relationships of the Galactic {\em disk} are
appropriate for the high-\z\ structures seen in \ngc 891.  As
discussed in \pI, the gas-to-dust relationships could be altered by
dust destruction in hydrodynamical processes (Jones \etal\ 1994) or by
the separation of the gas and dust through the effects of radiation
pressure (Ferrara \etal\ 1991; Davies \etal\ 1998).  Our images
suggest that the destruction of dust grains cannot be too severe,
given that high-\z\ dust is visible in any form. Though radiation
pressure could play a role in the shaping of the visible dust
structures, they are opaque enough to make such a mechanism relatively
inefficient.

The column densities given in Table \ref{table:derived} are roughly
consistent with $N_{\rm H} \sim \e{21}$ \column, or greater.  The
extinction values and column density estimates suggest that molecular
material may be present in these structures (e.g., Savage \etal\ 1977;
see also Elmegreen 1985 and McKee 1989 for theoretical viewpoints).
The dust features identified in \ngc 891 may contain large amounts of
mass, of order $\e{5} - \e{6}$ \msun, assuming Galactic gas to dust
relationships.  The masses are reminiscent of the Galactic giant
molecular clouds (GMCs; e.g., Solomon \& Sanders 1985).  The implied
potential energies are also quite large given these masses and large
heights above the plane.  Sandage (1961) also suggests that the
support of an extended, dusty layer of material in \ngc 891 must
entail large amounts of energy.

It should be pointed out that while detailed radiative transfer models
are beyond the scope of this paper, some of the more sophisticated
approaches to radiative transfer through a clumpy medium (e.g.,
Kuchinski \etal\ 1998; Witt \& Gordon 1996) could perhaps help answer
some of the more important questions regarding the properties of the
observed high-\z\ dust.  For example, detailed radiative transfer
models might help shed light on the total mass of the dust at high-\z\
and on the effects of high-\z\ dust opacity on optically-derived scale
heights (e.g., for determining the scale height of DIG emission).

\subsection{Notes on Individual Absorbing Structures}

Several of the absorbing structures identified in Figures \idfigs\
warrant further discussion.  We briefly comment here on a few of the
more spectacular structures visible in our images.

NGC 0891:D $-089+039$: As identified in Figure \ref{fig:IDsw} this
structure is an individual cloud at $z\sim1.75$ kpc.  However, there
are several individual clouds aligned vertically that may make up a
seemingly coherent structure traceable from $z\sim0.2$ to $\sim 2.0$
kpc.  Figure \ref{fig:filament1} shows a close up view of the unsharp
masked V-band image centered on the region around $-089+039$.  Several
very high-\z\ dust structures can be seen in this region.  If
$-089+039$ represents the upper-most clump of an irregular filament,
the low-\z\ anchor is a ring-like structure at $z\sim270$ pc whose
diameter is similar to the width of the filament near its base
($\sim210$ pc).  Our \Ha\ images, which will be discussed below, show
an \HII\ region coincident with the inside of the dust ring.  The
irregular absorbing clump in the center of the dust ring seems to lie
in front of the \HII\ region.  We find no evidence for \Ha\ emission
associated with the filament itself.  Like many of the dusty
structures seen in our images $-089+039$ and the lower-\z\ portions of
the filament are seen in absorption against the diffuse \Ha\ emission
of the ionized thick disk (see below).
 
Haffner, Reynolds, \& Tufte (1998) have found an ionized filament in
the DIG of the Milky Way that seems to stretch to $z\sim1.2$ kpc as a
coherent feature (as evidenced by its velocity structure).  However,
this filament is {\em only} seen in \Ha\ emission, with \HI, X-ray,
and IR emission being absent.  The suggested column density is $N_{\rm
H} \sim \e{19}$ \column, and the stringent limits on the \HI\ column
density suggest the structure is almost fully ionized.  The Haffner
\etal\ filament therefore has significantly different physical
conditions than the extended filament tentatively identified with
$-089+039$.

If the clouds visible beneath $-089+039$ are physically associated,
making a coherent, though irregular, filament more than 1.8 kpc in
length, it is a uniquely intriguing structure.

\begin{figure}
\epsscale{0.9}
\plotone{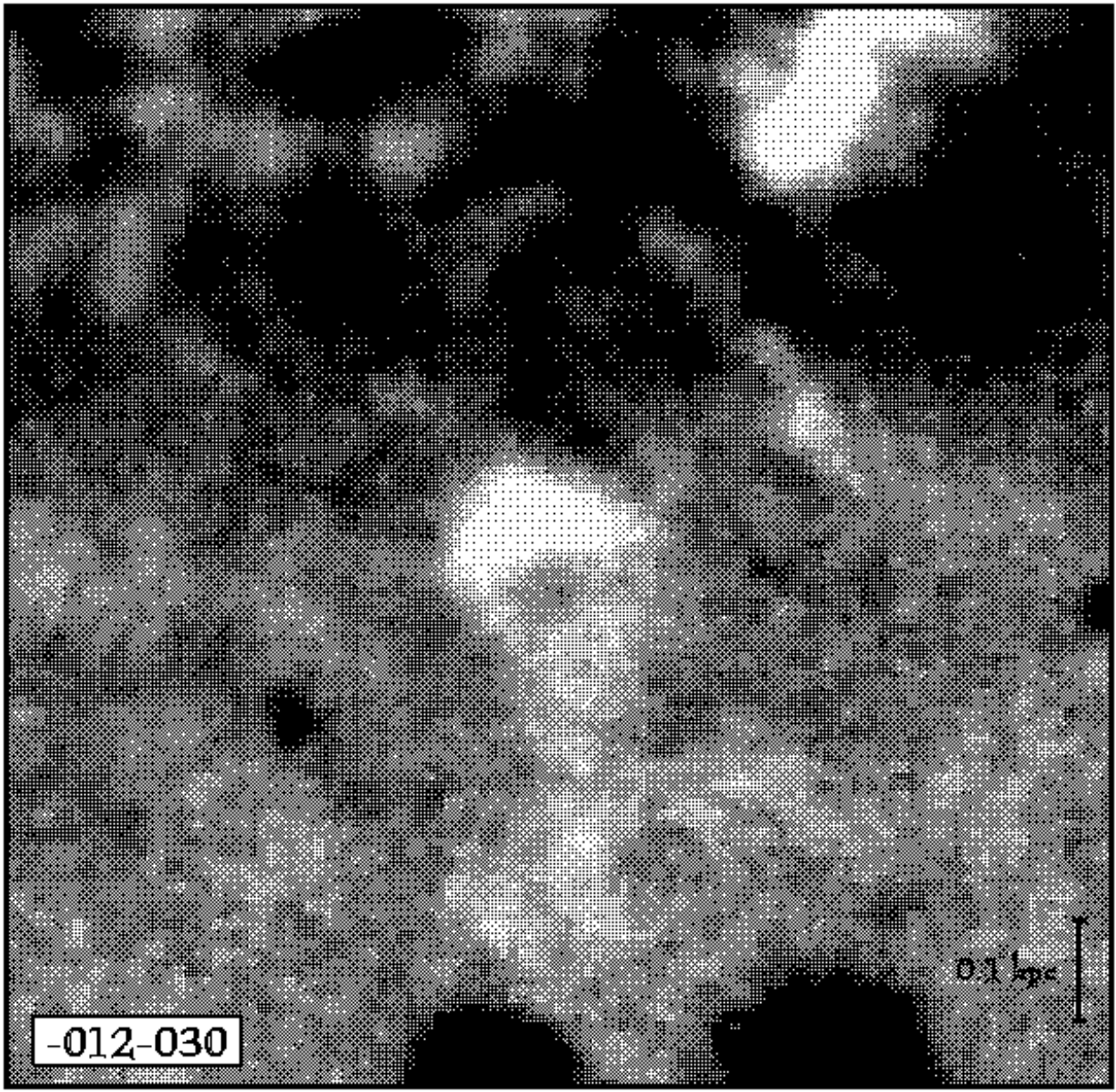}
\caption{A close-up view of the unsharp masked V-band data showing 
NGC~0891:D~$-012-030$.  This structure is a cometary-shaped cloud with
lower column density absorption trailing to high \z.  Only the
``head'' or low-\z\ component of this is described by the data
presented in Tables \ref{table:observed} and \ref{table:derived}.  The
$140\times 60$ pc head of this structure lies at a distance $z\sim1.3$
kpc from the midplane, though the trailing absorption can be traced
for an additional $\sim400$ pc.  The section of the image shown here
is 2\arcmin (or 1.08 kpc) on a side.  A scale bar of 0.1 kpc length is
given.
\label{fig:cometary1}}
\end{figure}

NGC 0891:D $-012-030$: This structure (identified as feature 7 in \pI)
is clearly cometary shaped with its apex pointed towards the center of
the galaxy and a tail of lower column density material extending to
high \z.  We show a close-up of this structure in Figure
\ref{fig:cometary1}.  The center of the head of the cometary structure 
is at $z\sim1.3$ kpc, but the trailing material can be traced to
almost $z\sim1.7$ kpc.  This structure is exemplary of the
vertically-oriented structures at high \z\ on the southern side of
\ngc 891.  We find no evidence for \Ha\ emission associated with 
this structure.

NGC 0891:D $+122-016$ \& $+134-019$: Together these two structures
(identified as features 11a and 11b in \pI) seem to trace the walls of
a supershell centered near $z\sim 800$ pc with a diameter of $\sim600$
pc (see Figure \ref{fig:IDne}).  As discussed in \pI\ and below (\S
\ref{subsec:comparison}), there is \Ha\ emission loosely associated
with the dusty shell walls.  The \Ha\ emission from this structure has
been discussed also in Pildis \etal\ (1994).  Our images show no
``cap'' to the structure in the dust distribution, suggesting a
supershell open to high \z.  The
\Ha\ images show a patch of diffuse emission at $z\sim800$ pc that
resides near the center of the shell, but this seems not to be the top
of the structure, given that the walls traced by the \Ha\ emission can
be found to larger \z -heights.  Perhaps this structure represents a
supershell that has experienced ``blowout.''  We will address this
issue in more detail below when we discuss the \Ha\ emission line
images.

NGC 0891:D $-010+037$: Though this structure is interesting for its
height ($z\sim1.75$ kpc), perhaps its most interesting attribute is
that it obscures a background spiral galaxy (see Figure
\ref{fig:IDmid}).  The southern edge of this spiral, centered at
$\alpha_{2000} = 2^h \, 22^m \, 29\fs6$, $\delta_{2000} = +42^\circ \,
20\arcmin \, 59\farcs0$, is completely obscured in the B-band.  While
the V-band shows slightly more of the low-\z\ end of this galaxy,
there still appears to be a significant amount of the galaxy that is
simply missing.  The analysis of the light from this galaxy could in
principle give us a better estimate of the opacity though a high-\z\
dust structure.  Unfortunately such an analysis is complicated by the
slightly irregular, asymmetric structure of the galaxy.  Furthermore,
an unresolved source is seen at the apex of the curve traced by the
dust structure ($\alpha_{2000} = 2^h \, 22^m \, 29\fs7$,
$\delta_{2000} = +42^\circ \, 20\arcmin \, 56\farcs6$), and directly
over the expected emission from the background spiral, in the I-band.
This source is not present in either the V- or B-band images.  The
archival \hst\ image presented in
\pI, which is taken through a very broad filter that encompasses much
of the V and R bandpasses, clearly shows an unresolved source as well.
This object could conceivably be an intermediate age or globular
cluster lying behind $-010+037$, and hence being extincted by this
dust structure.  It is likely not a supernova associated with the
background spiral given its relatively constant appearance in images
taken over several epochs.  If it is a star forming region in the
background galaxy, it is much brighter than any other such regions
that might exist in the galaxy.

It is clear from visual inspection that this background spiral is very
heavily extincted in the B and V images.

\section{The Relationship Between Dust and Ionized Gas in the 
Thick Disk of NGC~891}
\label{sec:halpha}

\subsection{Extraplanar Diffuse Ionized Gas}
\label{subsec:dig}

Emission from extraplanar DIG in \ngc 891 was first detected by
Dettmar (1990) and Rand \etal\ (1990) through narrowband imaging.
Spectroscopic studies of the DIG have more thoroughly characterized
its distribution and properties (Dettmar \& Shultz 1992; Keppel \etal\
1991; Pildis \etal\ 1994; Rand 1997, 1998).

The DIG layer has been fit by a two-component exponential distribution
including a thick disk with an electron scale height $h_z \sim 1$ kpc
and a more extended halo with a more uncertain scale height $h_z
\sim 2 - 5$ kpc (Hoopes \etal\ 1999; Rand 1997).  It should be noted
that these fits have made no correction for the dust that is clearly
visible in our broadband images.  Fits constrained to match only the
light in excess of $z \ga 1.0 - 1.5$ kpc should be reasonably secure.
Below $z \sim 1.0$ kpc, the covering factor of the dust features is
greater than unity, and below $z \sim 0.5$ there is virtually no
information on the intrinsic distribution of optical light from \ngc
891.

The power required to ionize the DIG can only be comfortably met by
the ionizing radiation from early type stars (e.g., Rand \etal\ 1990).
The strength of forbidden line emission from [\ion{N}{2}] and
[\ion{S}{2}] relative to \Ha\ increases with height above the plane
(Dettmar \& Shultz 1992; Rand 1997, 1998), as does the strength of
[\ion{O}{3}] to H$\beta$ (Rand 1998).  Pure photoionization models for
the DIG (e.g., Sembach \etal\ 2000; Domg\"{o}rgen \& Mathis 1994) with
very dilute ionizing radiation fields can match the high [\ion{N}{2}]
and [\ion{S}{2}] to \Ha\ ratios, though the very highest [\ion{N}{2}]
and [\ion{S}{2}] to \Ha\ ratios may require heating mechanisms beyond
those in low-density ionized nebulae (Reynolds, Haffner, \& Tufte
1999).  The increase in the [\ion{O}{3}]/H$\beta$ ratio suggests a
secondary source of ionization and/or heating may be at work (see
discussion in Rand 1998).  Given the very high [\ion{N}{2}] and
[\ion{S}{2}] to \Ha\ ratios observed in the extraplanar DIG, light
scattered from low-\z\ nebulae off of high-\z\ dust grains cannot
dominate the observed high-\z\ \Ha\ emission (see Ferrara \etal\ 1996;
Wood \& Reynolds 1999).

Figure \ref{fig:Hafull} shows our continuum-subtracted \Ha\ image of
\ngc 891.  The display in the top panel of Figure \ref{fig:Hafull}
shows the faintest diffuse gas, particularly the often faint high-\z\
material, while saturating the brightest emission.  The display in the
bottom panel is intended to show the structure of only the brightest
\Ha, including the very bright structures present above the plane in
the northeast section of the disk.  We comment briefly here on a few
of the important aspects of the emission line images presented in
Figure \ref{fig:Hafull}.  Our primary interest in the \Ha\ images is a
comparison of the extraplanar DIG and dust morphologies (see \S
\ref{subsec:comparison}).  Thus we will not detail the many aspects of
the ionized gas layer (see Dettmar 1990 and Rand \etal\ 1990 for
descriptions of the DIG distribution).  However, we will briefly
discuss the properties of several individual ionized filaments in \S
\ref{subsec:filaments}.

Our \Ha\ images clearly show emission from the DIG at large distances
from the midplane.  We detect \Ha\ emission to heights exceeding
$z\sim4$ kpc, though our sky subtraction difficulties do not allow us
to accurately characterize the highest-\z\ emission.  Some of the \Ha\
emission is in the form of discrete large-scale filaments, most of
which seem to run roughly perpendicular to the midplane.  There is
also a more diffuse background of \Ha\ emission visible along most of
the disk and to high \z.  Several large \HII\ regions on the front
side of the galaxy dominate the light near the midplane.  As noted by
Rand \etal\ (1990) and Dettmar (1990), there is a strong asymmetry in
the strength of the emission between the northeast and southwest sides
of the disk (left and right, respectively, in our images); this
asymmetry is also seen in the radio continuum observations of Dahlem,
Dettmar, \& Hummel (1994) and in the far infrared (50 $\mu$m)
observations of Wainscoat, de Jong, \& Wesselius (1987).  We also see
an asymmetry in the light distribution in our broadband images, with
the northeast section of the disk being brighter than the southwest
section at most heights $z\la1.0 - 1.5$ kpc.  This appears to be due
to a greater amount of intervening high-\z\ absorbing material on the
southwest side of \ngc 891.  The \Ha\ asymmetry is much stronger than
that seen in the broadband images, and the radio continuum
measurements are not affected by dust, implying a real asymmetry in
the amount of star formation between the two sides of \ngc 891.

Aside from the known extraplanar DIG, we also detect what appear to be
several discrete \HII\ regions at high \z\ in \ngc 891.  These
candidate nebulae are visible in Figures \ref{fig:HaDustMid}, 
\ref{fig:HaDustNE}, and \ref{fig:HaDustSW} which show the middle, 
northeast, and southwest sections of the disk, respectively, in \Ha\
and dust.  These figures will be discussed in more detail below.  In
Figure \ref{fig:HaDustMid} an unresolved source in the \Ha\ image is
seen in the northern side of the disk (in the upper right corner of
this display) corresponding to an unresolved continuum source in the
broadband images.  This object lies at $z\sim1.3$ kpc above the
midplane.  A similar object is seen in Figure
\ref{fig:HaDustNE}, also on the northern side of the disk (above the
disk, near the middle in this display).  This object is again
coincident with an unresolved source in the broadband images at
$z\sim0.6$ kpc.  Several such unresolved knots of \Ha\ emission can be
seen in Figure \ref{fig:HaDustSW} at heights $z\sim900-2000$ pc, all
of which are coincident with continuum sources.  These objects are not
among the list of the planetary nebulae identified in \ngc 891 by
Ciardullo, Jacoby, \& Harris (1991).  These candidate \HII\ regions
and their associated continuum sources will be discussed in a future
publication (Howk \& Savage, in preparation).

\subsection{Large-scale Filamentary Structures in the Ionized Gas}
\label{subsec:filaments}

In this section we very briefly summarize the properties of several of
the filamentary structures observed atop the more diffuse component of
the DIG in \ngc 891.  We do this both because it allows the reader to
compare the general properties and distribution of the ionized gas
filaments with those traced by the dust, and because no previous
catalog of such structures exists in the literature, though there
seems to be a general consensus that this filamentary component is
important (Rand 1997, 1998; Rand \etal\ 1990; Dettmar 1990).

\begin{figure}
\epsscale{0.9}
\plotone{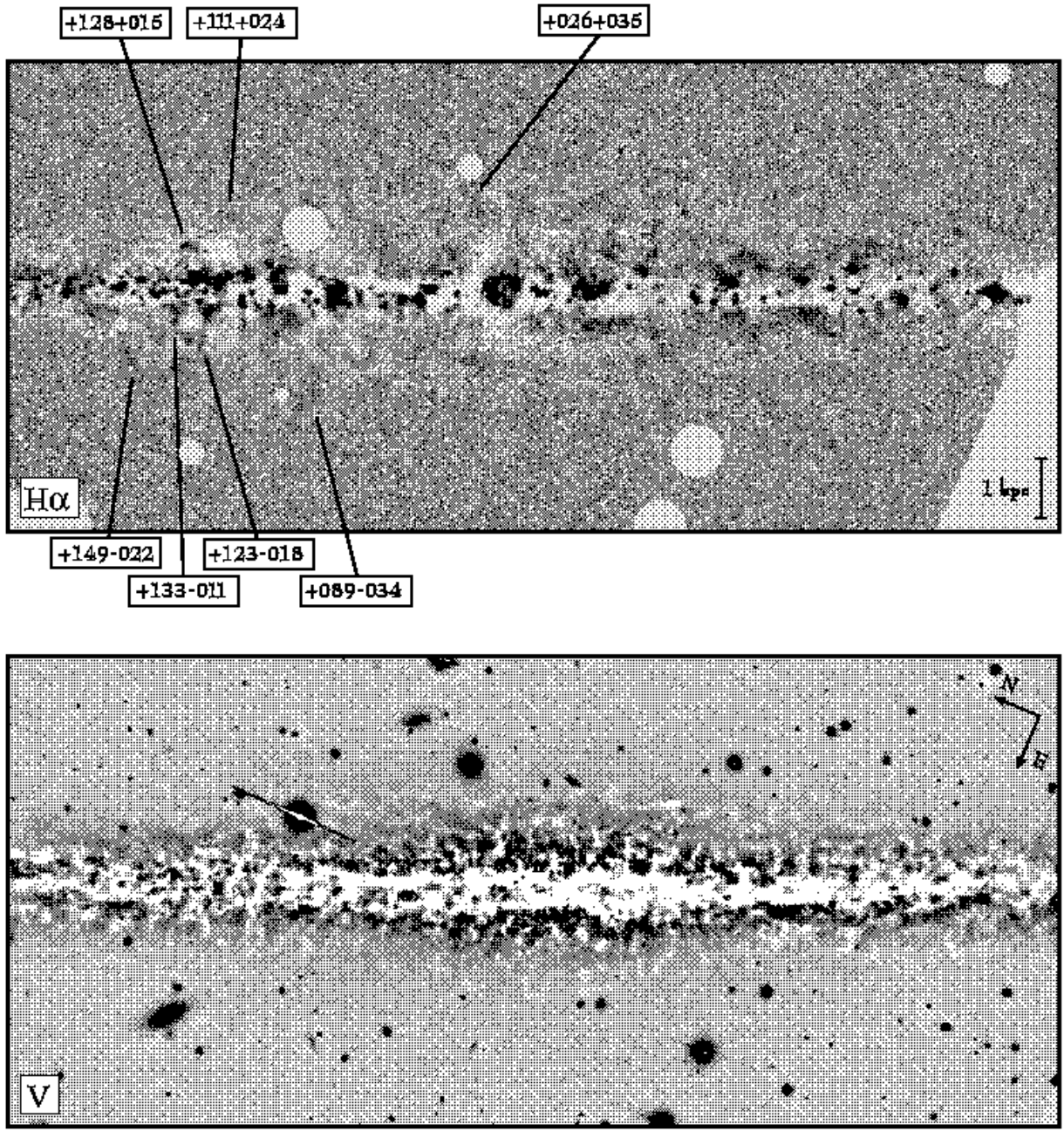}
\caption{Unsharp masked views of both the continuum-subtracted
\protect\Ha\ emission line image of \ngc 891 (top) and the V-band
image (bottom).  This image covers $6\farcm4 \times 2\farcm8$ ($17.3 \
{\rm kpc} \, \times7.6$ kpc).  North and east are marked, and a 1 kpc
scale bar is shown in the lower right portion of the \protect\Ha\
image.  The unsharp masked images (in both bands) were produced by
dividing the original images by the images smoothed with a Gaussian
kernel with FWHM$\, = 35$ pixels ($6\farcs9$). Several individual DIG
filaments are identified in the upper panel.  The properties of these
structures are summarized in Table \ref{table:filaments}.  It is clear
from these images that there is significantly less structure in the
\Ha\ images, which trace the DIG, than in the V-band images, which
trace the dust absorption.
\label{fig:filaments}}
\end{figure}

Figure \ref{fig:filaments} shows another view of our high-resolution
\Ha\ observations of \ngc 891.  We have applied an unsharp masking 
procedure to the \Ha\ data displayed in the top panel of Figure
\ref{fig:filaments}, while the bottom panel shows the V-band unsharp 
masked image from Figure \ref{fig:Vfull} for comparison.  To produce
the unsharp masked \Ha\ image shown in Figure \ref{fig:filaments}, we
used a Gaussian smoothing kernel with FWHM$\, = 35$ pixels
($6\farcs9$), the same used in producing the V-band unsharp masked
images.

We have identified and labelled a small number of filamentary
structures in Figure \ref{fig:filaments}.  We summarize the rough
properties of the identified filaments in Table \ref{table:filaments}.
These properties include coordinates, \z -heights, and dimensions.  We
use a naming scheme like that adopted for the dust structures in \S
\ref{subsec:nomenclature}, identifying each filament using the form
\ngc 0891:DIG $\pm$XXX$\pm$ZZZ.  As with our discussion of individual 
dust structures, our selection of \Ha -emitting features is biased
towards the more spectacular examples at high \z\ (where the confusing
influence of dust absorption is lessened).

Figure \ref{fig:filaments} shows much less small-scale structure in
the \Ha\ unsharp-masked image than the unsharp-masked images of the
V-band data, which accentuate the absorption by dust.  In part the
smaller number of filaments in the \Ha\ images is caused by the
sensitivity of our observations to interstellar matter: where the
dusty filaments are present in galaxies, they are much easier to
observe than the low-intensity \Ha -emitting filaments.  However, it
is clear from Figures \ref{fig:Hafull} and \ref{fig:filaments} that
the filamentary structures in the DIG do not show as great a contrast
compared with the background emission as the dust structures, which
are visible only because they are much more opaque than their
surroundings.

Not only are the \Ha -emitting filaments less numerous (and less
prominent) than the dust structures seen in absorption, but much of
the \Ha\ emission seems to be in the form of a more
smoothly-distributed medium.  However, the appearance of this
``smooth'' component is made more complicated by the absorption caused
by the dust features seen in our broadband images.  We have avoided
identifying any apparent structures that may be caused by absorption
due to intervening dust.

The distribution of dust-laden filaments is more symmetric about the
center of \ngc 891 than the distribution of \Ha\ filaments.  For
example, Figure \ref{fig:Vsw} shows the southwest region of the disk
contains a number of prominent dust features.  Though a few \Ha
-emitting structures seem to be visible in this region most of them
are the result of absorption by the irregularly-distributed
dust-bearing clouds.  The relatively small number of \Ha\ filaments in
the southwest region of the disk compared with the northeast region is
consistent with the general asymmetry in the \Ha\ emission across the
center of \ngc 891 (Rand \etal\ 1990; Dettmar 1990; Dahlem \etal\
1994).

In general the filaments we have identified in Figure
\ref{fig:filaments} are oriented roughly perpendicular to the plane of
\ngc 891.  Most are singular arcs of emitting material, though a few
are in the form of shell-like structures (e.g., \ngc 0891:DIG
$+089-034$ and the pair $+123-018$ and $+133-011$).  The faint ionized
structure \ngc 0891:DIG $+089-034$ is one of the more spectacular
features in our \Ha\ image.  As identified in Figure
\ref{fig:filaments} (see also Figure \ref{fig:Hafull}), it forms a 
bubble-like structure approximately 650 pc in diameter, centered
$\sim1500$ pc above the midplane.  This structure is closed on itself,
and if it traces a superbubble there is no evidence for break-out.
However, it is interesting as much for its location above the plane of
the galaxy as its morphology.  Even single supernova explosions in
low-density halos of galaxies can effect their environments on very
large scales (e.g., Shelton 1998), and this structure could
conceivably be the result of a high-\z\ supernova.

The shell traced out by the pair of filaments \ngc 0891:DIG $+123-018$
and $+133-011$, which are associated with the dust structures \ngc
0891:D $+122-016$ and $+134-019$, respectively (see below), may also
be tracing a large ($\sim600$ pc diameter) shell in the ISM, though at
lower \z.  This shell, centered at $z \approx 800$ pc, has no obvious
``cap'' or top, though there is a diffuse patch of emission
immediately between these filamentary shell walls.  These ionized
filaments, along with \ngc 0891:DIG $+128+015$ are discussed in some
detail by Pildis \etal\ (1994).

A few of the \Ha -emitting filaments seem to be more diffuse (thicker)
than the typical dust structures (e.g., \ngc 0891:DIG $+089-034$ and
$+026+035$), though most of those listed in Table \ref{table:filaments}
are of comparable size to the absorbing dust clouds.  The emission
measures towards the brightest filaments identified in Table
\ref{table:filaments} are 2--3 times that of the nearby diffuse component
of the \Ha\ emission, while the fainter are only $30\% - 50\%$
brighter.  The brightest filaments are only present above the region
of vigorous star formation in the NE section of the disk.

There is no evidence in any of our images for absorption that is
directly associated with the filaments observed in our \Ha\ images.
This does not imply these structures lack dust.  If we assume that the
filaments have emission measures approximately the same as the average
at $z\approx 1$ kpc, i.e., $EM\approx10$ pc cm$^{-6}$ (e.g., Rand
\etal\ 1990), and assume densities of $n_e
\approx 0.2$ \percc, then the hydrogen column densities in these
filaments should be $N_{\rm H} \sim \e{20}$ \column.  If the
gas-to-dust ratio in the filaments is similar to gas in the Galactic
disk, these structures should not show detectable absorption in our
images.  We are not sensitive to extinction in the more diffuse
component given our method of identifying dust in our images.  We
discuss in detail the relationship between the observed dust-laden
filaments and the morphology of the \Ha\ emission in the next
subsection.

\subsection{The Distribution of Extraplanar Ionized Gas and Dust}
\label{subsec:comparison}

The \Ha\ images presented in Figure \ref{fig:Hafull} show that the
extraplanar DIG in \ngc 891 contains many filamentary structures, some
stretching to quite large \z .  However, comparing the \Ha\
distribution with that of the dust seen in Figure \ref{fig:Vfull}
suggests that the \Ha-emitting material is more smoothly distributed
than the matter traced by the dust absorption.  Are these two views of
the high-\z\ ISM in \ngc 891 tracing the same structures?  The answer
to this question is almost certainly no.  There are very few regions
in the thick disk of this galaxy where the DIG and dust structures are
associated with one another in our images, and in those few regions
the association is only a loose one.  There are a large number of
regions where the \Ha\ emission is obscured by the dust features,
usually in much the same amount as the stellar continuum emission seen
in our broadband images.  Studies of \Ha\ emission from \ngc 891 must
systematically underestimate the ionized gas emission measures at
heights $z \la 2$ kpc, though the effect is most important for $z
\la 1$ kpc.  The extinction due to the dust may also cause the observed
distribution of light to differ from the intrinsic distribution;
emission scale heights derived from fits to $z \la 1$ kpc may need
to be reconsidered given the observed absorption of the \Ha\ emission
by the dust structures in our images.

\begin{figure}
\epsscale{0.95}
\plotone{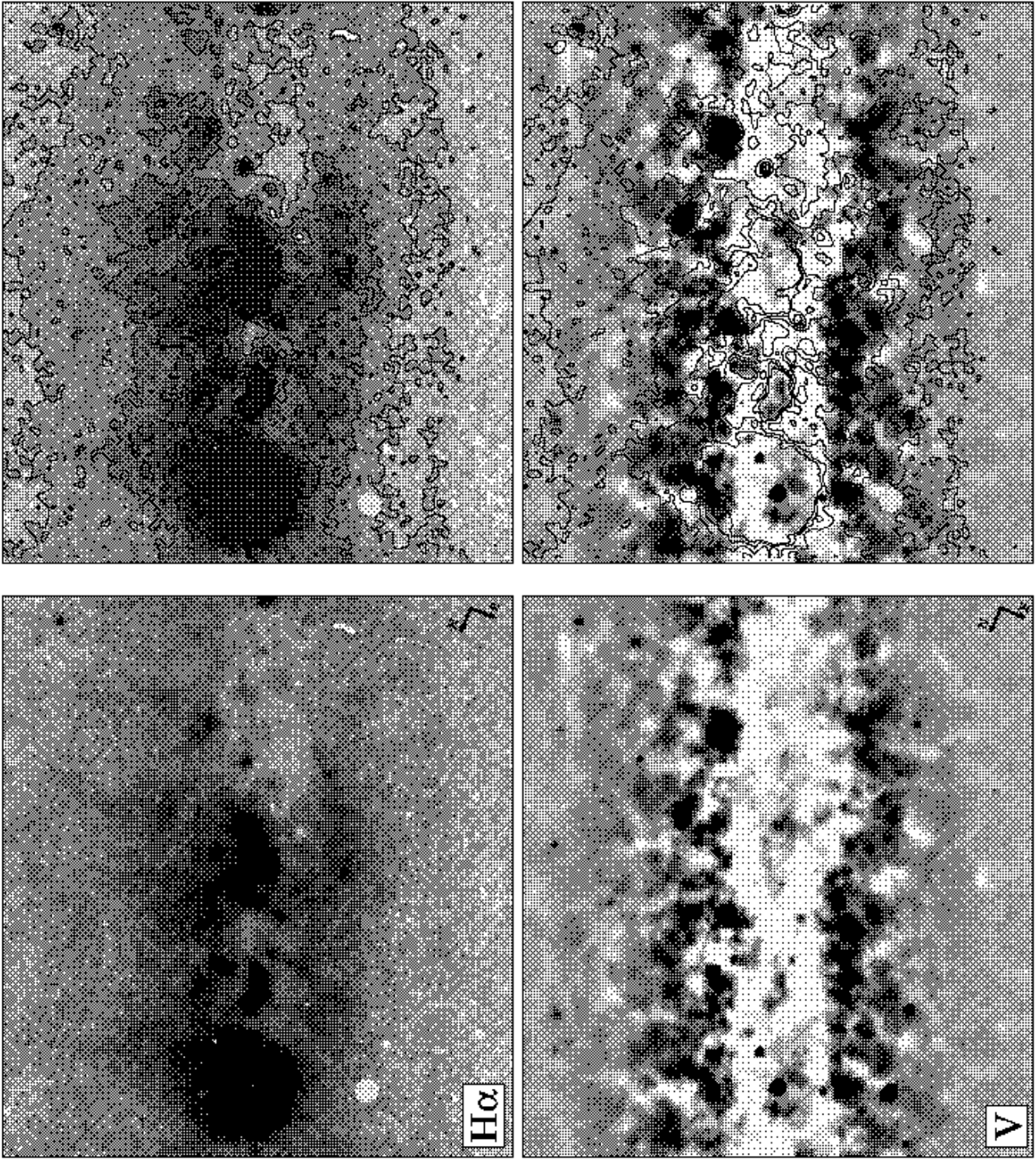}
\caption{A comparison of the \Ha\ emission and dust
absorption near the center of \ngc 891.  The continuum-subtracted \Ha\
and V-band unsharp masked data for this region are shown as grayscale
in the upper and lower panels, respectively.  The two right hand
panels have \Ha\ contours overlaid on the grayscale images.  The \Ha\
grayscale is shown using a logarithmic stretch.  The \Ha\ contours are
approximately $5\sigma, \, 10\sigma, \, 15\sigma, \, 20\sigma,\ {\rm
and} \ 25\sigma$ above the background.  The region shown in this image
covers $1\farcm3 \times 1\farcm2$ ($3.5 \ {\rm kpc} \, \times3.2$ kpc)
and is centered $\sim0.5$ kpc southwest of the galaxy center.  Small
bright white spots in the \Ha\ grayscale are due to the effects of
poorly removed cosmic rays in the off-band
images. \label{fig:HaDustMid}}
\end{figure}

\begin{figure}
\epsscale{0.95}
\plotone{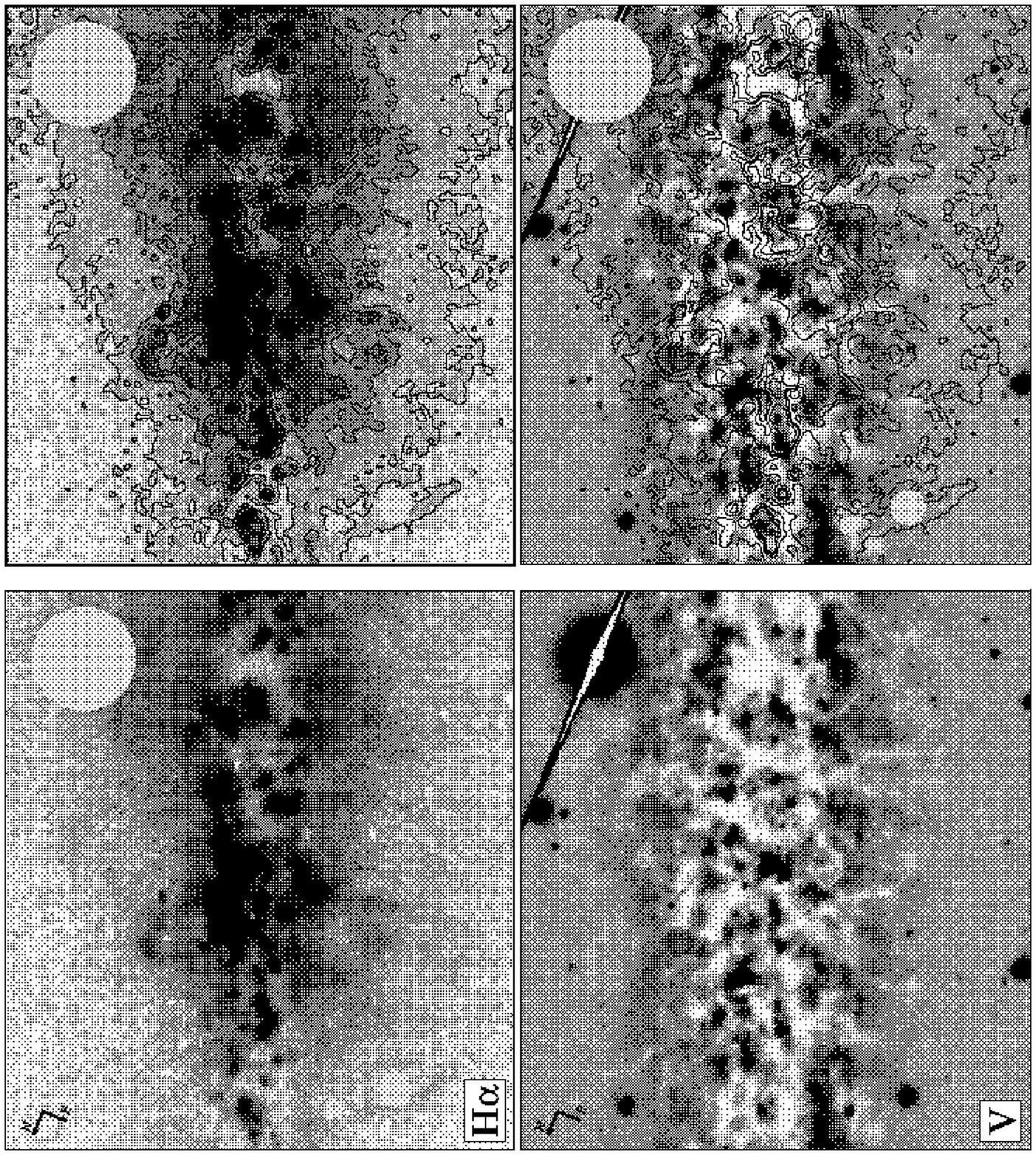}
\caption{As Figure \ref{fig:HaDustMid} but for a section of the disk centered
$\sim5.5$ kpc northeast of the nucleus of \ngc 891.
\label{fig:HaDustNE}}
\end{figure}

\begin{figure}
\epsscale{0.95}
\plotone{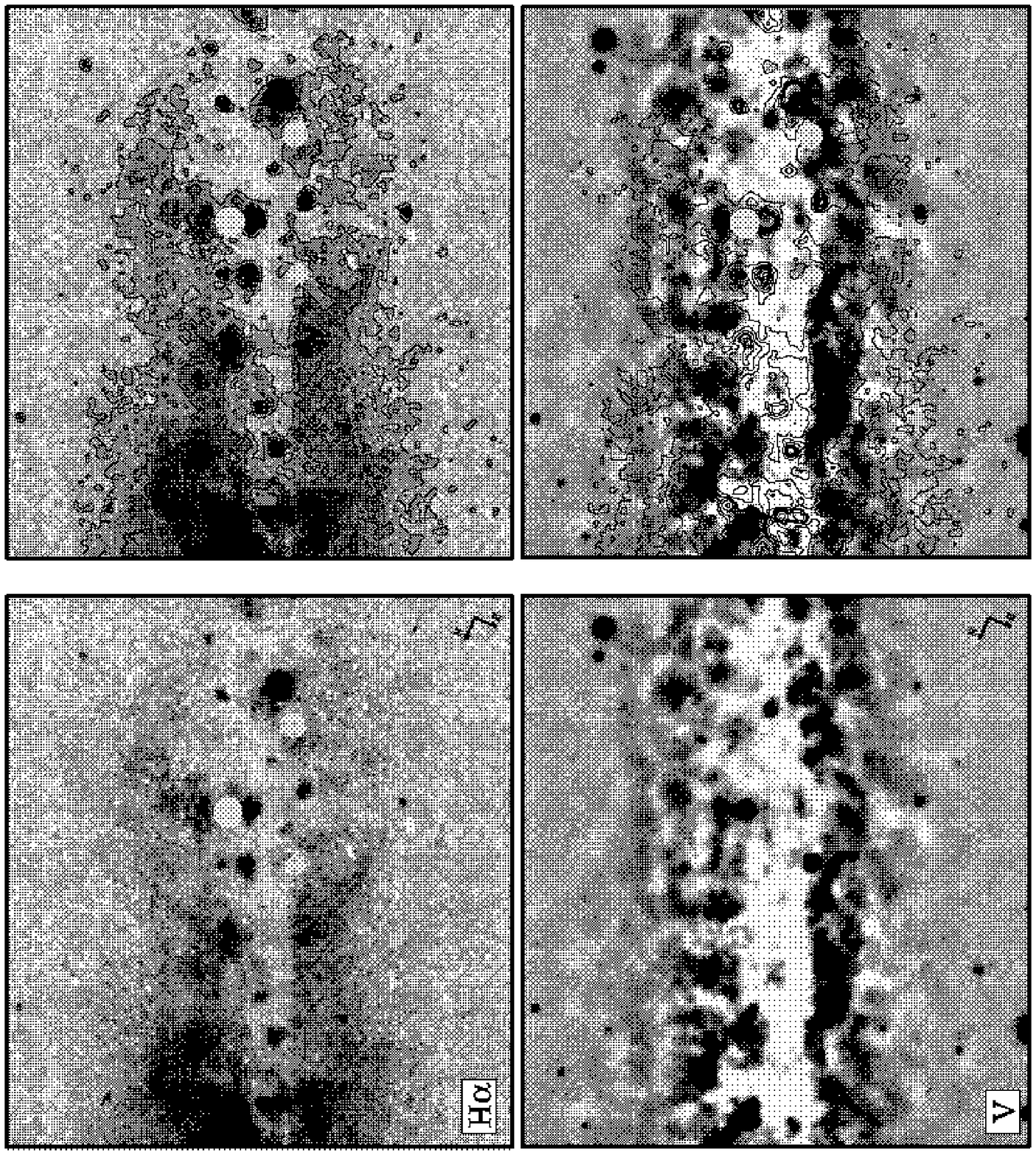}
\caption{As Figure \ref{fig:HaDustMid} but for a section of the disk centered
$\sim4.4$ kpc southwest of the nucleus of \ngc 891.
\label{fig:HaDustSW}}
\end{figure}

Figures \ref{fig:HaDustMid}, \ref{fig:HaDustNE}, and
\ref{fig:HaDustSW} show a comparison of the dust and \Ha\ morphology
for three regions, one near the center of the galaxy (Figure
\ref{fig:HaDustMid}), one $5.5$ kpc projected distance
northeast of the center (Figure \ref{fig:HaDustNE}), and the last
centered $4.4$ kpc to the southwest of the center (Figure
\ref{fig:HaDustSW}).  Each of these figures shows grayscale images of
the ionized gas emission as well as the V-band unsharp masked view
from a $1\farcm3 \times 1\farcm2$ ($3.5 \ {\rm kpc} \, \times3.2$ kpc)
region of \ngc 891 (top and bottom two panels, respectively).  In the
right two panels, contours of the \Ha\ emission are overlaid on top of
these grayscale images.  The \Ha\ contours can be slightly misleading
given they show a small number of quantized levels of emission.  They
do, however, offer a reasonable way of displaying one dataset on top
of another.  One should use both the grayscales and contours for
comparing in detail the distribution of dust and ionized gas and dust
in these figures.

It is clear from Figures \ref{fig:HaDustMid},
\ref{fig:HaDustNE}, and  \ref{fig:HaDustSW} that any direct association 
of the extraplanar ionized gas and dust at high \z\ is very weak.
Towards the center of the galaxy, shown in Figure \ref{fig:HaDustMid},
a significant amount of the structure seen in the \Ha\ images is
simply caused by the absorption of \Ha\ photons by foreground dust
features.  The \Ha\ contours are often just offset from the dust
features.  But the general impression is not one of \Ha\ brightening
at the edges of the dust features, but rather the dimunition of the
more smoothly distributed background emission.  The southwest section
of the disk displayed in Figure \ref{fig:HaDustSW} gives a very
similar impression.  In particular one can quite clearly see evidence
for absorption of \Ha\ photons by the low-\z\ portions of the dust
structures $-089+039$ and $-107-020$ (see Figure \ref{fig:IDsw}).

One-dimensional cuts of the light distribution of the \Ha\ emission
reinforce this conclusion that the \Ha\ light is being diminished by
the absorbing dust structures.  Typical measurements of the apparent
extinction values towards a number of dust features in our \Ha\ images
yield results intermediate between the $a_V$ and $a_I$ values.  There
is typically no increase in the brightness of the \Ha\ emission as one
nears the location of a dust feature, but rather a dimunition of the
light from ionized gas.

There are certainly filaments in the \Ha\ images that are not simply
caused by dust absorption effects, some of which are discussed in the
previous subsection.  Some of the filaments have also been observed by
Rand (1998) to have slightly different forbidden emission line
intensities relative to \Ha.\footnote{Rand's slit for these
observations was oriented parallel to the plane at a height
$z\approx700$ pc above the midplane.} This suggests these structures
may be physically distinct from the general background emission and
hence not simply the result of dust absorption.

\begin{figure}
\epsscale{0.75}
\plotone{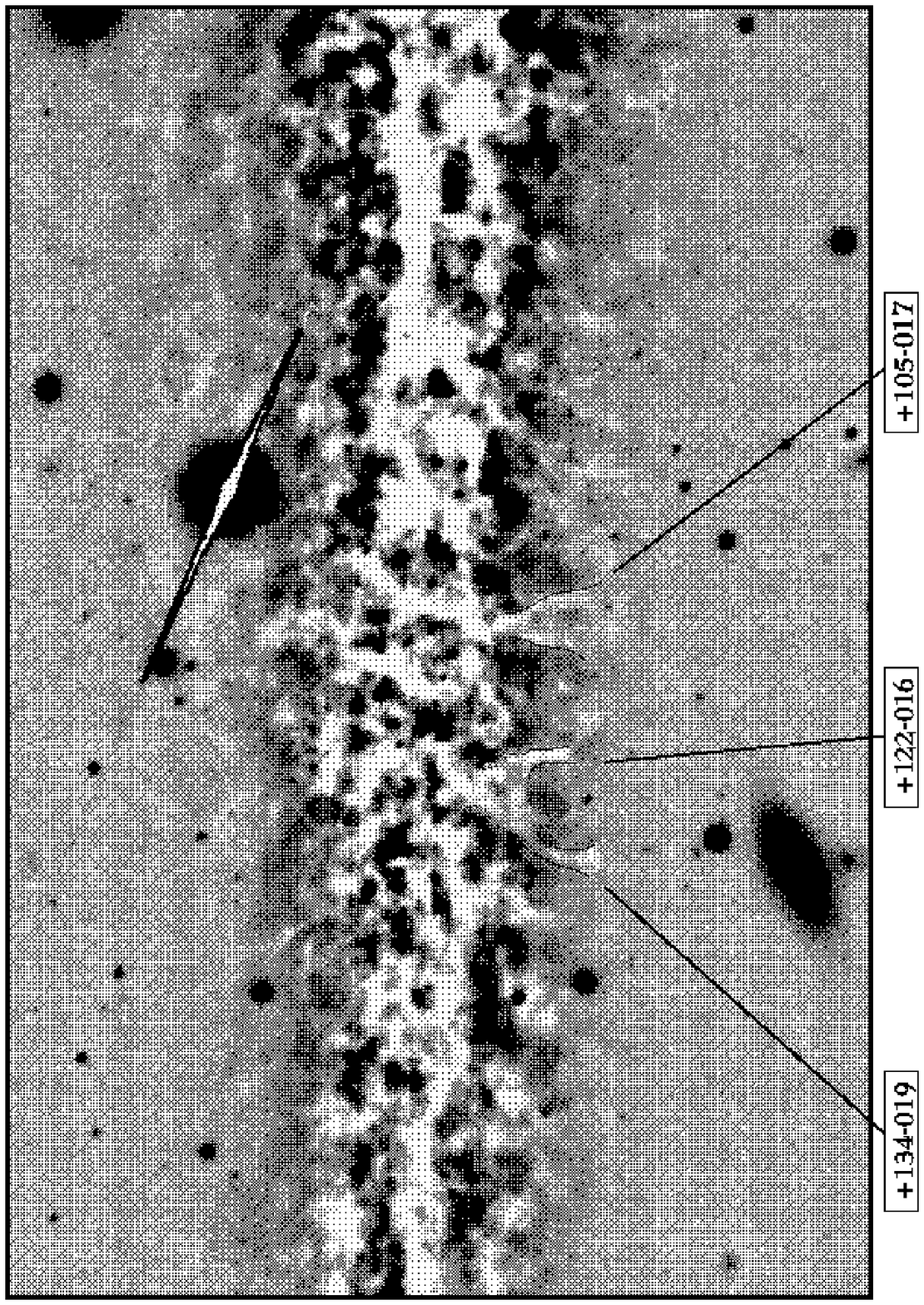}
\caption{The same region as Figure
\ref{fig:Vne}, but with several individual absorbing structures
identified.  The properties of these clouds are summarized in Tables
\ref{table:observed} and \ref{table:derived}.
\label{fig:IDne}}
\end{figure}

\begin{figure}
\epsscale{1.0}
\plotone{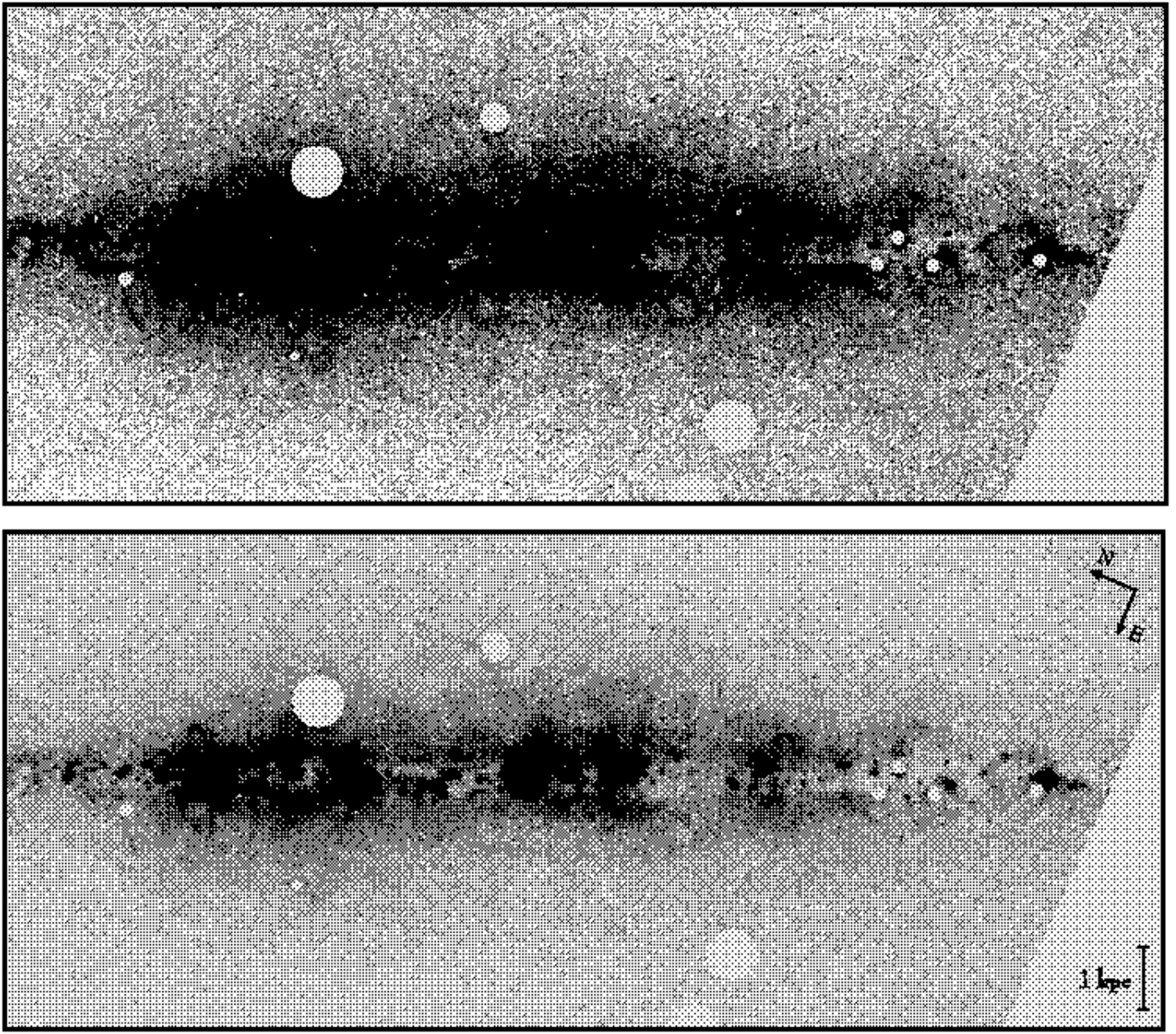}
\caption{The continuum-subtracted \Ha\ emission line image of \ngc
891.  The image is displayed in the top panel to show the faintest
diffuse gas, while the display in the bottom panel shows the
distribution of bright \Ha\ emission.  These images cover the same
area, $6\farcm4 \times 2\farcm8$ ($17.3 \ {\rm kpc} \, \times7.6$
kpc), and have the same orientation as those shown in Figure
\protect\ref{fig:Vfull}.  A bar denoting 1 kpc is given for scale in
the lower right hand corner of the lower panel.  We have placed gray
circles over regions where the halo of the point spread function from
bright stars in our on-band images may contaminate the view of the
ionized gas emission.  Note that the \Ha\ emission is much less
structured than the dust absorption illustrated in Figures
\ref{fig:Vfull}--\ref{fig:cometary1}.
\label{fig:Hafull}}
\end{figure}

In the northeast section of the disk, shown in Figure
\ref{fig:HaDustNE}, we see several dust structures that appear to have
been shaped by the effects of star formation (e.g., $+105-017$,
$+122-016$, and $+134-019$; see Figure \ref{fig:IDne}).  This region
of the disk was imaged by Pildis \etal\ (1994), who described the \Ha\
emission from two ionized ``supershells.'' This side of the disk of
\ngc 891 is experiencing vigorous  star formation as traced by various means 
(e.g,. Rand \etal\ 1990; Dettmar 1990; Dahlem \etal\ 1994).

A few of the dust structures in the northeast section of the disk,
particularly those with morphologies suggestive of shells and cones,
show a loose association of the DIG emission and the dust absorption
structures; an association that can likely not be caused by the simple
attenuation of background emission.  Examining Figure
\ref{fig:HaDustNE} we see the twin walls of a shell traced by the dust
structures $+122-016$ and $+134-019$ on the southern side of the plane
(below the plane in this figure).  The region immediately interior of
the shell walls traced by these dust features is bright in \Ha\
emission.  However, if one examines the position of the \Ha\ contours
with respect to the dust absorption, the alignment of the walls traced
by the dust absorption and the \Ha\ emission is not perfect.  In
particular the northern-most wall visible in absorption ($+134-019$)
is not particularly well aligned with the corresponding \Ha\ arc, and
may even be present in absorption against the \Ha\ emission at high
\z.  Interestingly, though there is an irregular clump of \Ha
-emitting material near the center of this shell, there is little
evidence for a top or cap to this shell in the \Ha\ images.

Another structure showing a loose association between the \Ha\
emitting material and that traced by the dust absorption is the
feature identified as $+105-017$ in Figure \ref{fig:IDne}.  This
structure, appearing as a vertical cone open to high-\z\ in the dust
absorption, appears to be edge-brightened in \Ha.  It is seen just to
the right of center, and below the plane, in Figure
\ref{fig:HaDustNE}. Both the dust absorption and associated \Ha\
emission can be traced to $z\sim0.8$ kpc from the midplane.

Though a few of the dust structures seen in Figure \ref{fig:HaDustNE}
show evidence for loosely associated \Ha\ emission, many of the dust
features have no clear-cut association with the extraplanar DIG.
Several of the high-\z\ absorbing features seen in Figure
\ref{fig:HaDustNE} seem to be attenuating the \Ha\ emission.  This
attenuation of the \Ha\ emission by foreground dust structures is seen
along much of the length of the galaxy.  The dust is seen to be much
more highly clumped than the DIG emission in our images.  When viewing
our \Ha\ images in detail, one is left with the impression that the
DIG emission may in fact be quite smooth, with a few filaments within
this smooth layer.

All of this discussion regarding an absence of a direct correlation
between the observed dust features and the DIG traced by the \Ha\
emission does not imply the DIG of \ngc 891 is free of dust (see \S
\ref{subsec:filaments}).  In our own Galaxy, there is increasing
evidence for dust in the warm ionized medium (Lagache \etal\ 1999,
2000; Howk \& Savage 1999a).  However, our broadband absorption
detection method would not be sensitive to the expected column
densities and smoother structure of the DIG in \ngc 891.

\section{The Multiphase Halo of NGC~891}
\label{sec:discussion}

The deep images presented in this work have shown that \Ha\ emission
from the extraplanar DIG in \ngc 891 is not generally correlated with
the dust-bearing clouds seen in absorption against the background
stellar light, although there is a very strong statistical correlation
between the presence or absence of these two phases of the ISM in
spiral galaxies (\pII).  Our observations also show the
highly-structured, or clumped, dusty clouds are prevalent only at
heights $z\la 2$ kpc from the midplane of \ngc 891 while \Ha\ emission
is seen to much larger \z\ distances.  Many of these dust clouds have
column densities in excess of $\e{21}$ \column, and enclose masses
$>\e{5}$ \msun.  We estimate the {\em average} densities of these
high-\z\ clouds are roughly $n_{\rm H} > 2-10$
\percc.

The lack of association between the \Ha\ emission from high-\z\ DIG
and the highly-structured dusty clouds seen in our broadband images
implies that the \Ha\ emission and the dust absorption are not tracing
the same interstellar material.  Our images suggest an arrangement of
the high-\z\ ISM in \ngc 891 where dusty dense clouds, visible through
their absorption of the background stellar light, coexist with a more
diffuse ionized medium.

We suggest that all of these aspects of the high-\z\ dust features can
be understood in the context of a multiphase ISM in the thick disk of
\ngc 891, possibly in a quasi-stable configuration.  The dusty clouds
seen far above the disks of spirals likely represent a dense phase of
the ISM at high-\z.  In this case they may be the result of thermal
instabilities in the high-\z\ ISM and may perhaps be in pressure
equilibrium with either the DIG or hot, X-ray emitting gas (e.g.,
Wolfire \etal\ 1995a,b).  Our images show the ionized gas is certainly
more smoothly distributed than the material traced by the dust clouds.
This suggests the volume filling factor of the dense phase is lower
than that of the DIG, in accord with theoretical expectations (e.g.,
McKee \& Ostriker 1977).  However, given the lack of association
between the DIG and the dense material seen in absorption, it is clear
that some portion of the warm ionized medium is not associated with
cool cores of neutral material, e.g., as envisioned by McKee \&
Ostriker (1977).

In \pII\ the multiphase nature of the ISM in the thick disks of many
nearby spiral galaxies was discussed in the context of a quasi-stable
configuration of several thermal ``phases.''  The calculations of the
allowable equilibrium thermal states of a multiphase ISM by Wolfire
\etal\ (1995a,b) were used as guidelines for estimating the requisite
conditions for such a configuration. These authors calculate detailed
thermal and ionization balance of a multiphase medium (see also McKee
\& Ostriker 1977; Field \etal\ 1969) and give the results of these 
calculations for a range of physical conditions of the medium.  A
minimum pressure requirement must be met to support a stable dense
phase of the ISM.  In the Wolfire \etal\ (1995a) models this pressure
for high column density clouds ($N_{\rm H} > \e{20}$
\column) with conditions appropriate for the disk of the Milky Way is
$P^{min}/k \approx 600$ K \percc.  Larger column densities will tend
to lower the required pressure, as will a decrease in the ambient
radiation field or in the dust to gas ratio relative to those
appropriate for the Milky Way disk.  Thus the minimum pressure
required to support the high-\z\ dust clouds in \ngc 891 could be
slightly lower than this value.  

If we assume the observed structures are cylindrically symmetric, so
that their depth along the line of sight is similar to their minor
axis lengths, the column density estimates given in Table
\ref{table:derived} suggest  these structures have densities
$\langle n_{\rm H} \rangle \ga 2 - 10$ \percc.  This is much greater
than the expected density of the DIG at {\em any} distance from the
plane, further suggesting that the \Ha\ emitting structures are
tracing a separate medium than that traced by the dust.  This gross
density estimate is consistent with the average densities predicted
for dense clouds in the Wolfire \etal\ (1995a) models (see also \pII).

Our rough calculations in \pII\ suggested the average pressures in the
DIG or hot ionized medium traced by X-rays were sufficient to support
a dense phase of the ISM.  For example, the average pressure provided
by the DIG is $P^{min}/k \approx 600$ K \percc\ at $z\sim 1.6$ kpc
(assuming the electron distribution of Rand 1997 and $T_e \approx
8,000$ K).  The X-ray emitting gas likely has a larger pressure
(Bregman \& Houck 1997).  The paucity of clouds beyond $z\sim2$ kpc is
understandable in this scenario: at heights in excess of $\sim2$ kpc,
the pressure of the ambient medium may be too low to support a dense
phase of the ISM.  In this case dust may still exist at larger heights
above the midplane, but our images are not sensitive to a diffuse
component of dust.  We are only able to see those regions that are
denser (i.e., more opaque) than their surroundings.  Given the
uncertainties in the distribution, temperature, filling factor, and
ionization fraction of the ambient gas, we believe the pressure
requirements for confinement of a dense medium are easily met in
\ngc 891 at heights $z \la 2.0$ kpc.

Other theoretical and observational lines of evidence support our
conclusion that the dust structures trace a dense medium at high-\z.
The three-dimensional numerical simulations of gaseous disk dynamics
by de Avillez (1999) suggest that the formation of dense interstellar
clouds at great \z -heights may be a relatively common by-product of
the circulation of material between the disk and halo of a galaxy.
His simulations follow the temperature, density, and dynamics of
parcels of gas in a model galactic disk undergoing star formation, and
hence experiencing energy input by supernovae.  The combined effects
of multiple supernovae produce a fountain-like flow in these models,
with gas continually flowing up from the disk and raining down upon it
from above.  These detailed simulations show cloud formation at
high-\z.  In these simulations the bulk of the clouds are formed
between $0.3 \la z \la 1.5$ kpc, and de Avillez suggests the
formation of dense clouds is much more efficient in regions compressed
by shock waves.  In particular he finds the intersections of shocks at
high-\z\ can serve to compress the gas, which then cools rapidly.

\garciaburillo, Combes, \& Neri (1999) have observed CO emission from
the edge-on galaxy \ngc 4013 at high angular resolution ($\sim 3
\arcsec$).  They find about 10\% of the emission in their maps comes
from high-\z\ material.  Our own WIYN images of this galaxy (\pII)
show a high-\z\ dust distribution similar to that seen in \ngc 891.
Several of the high-\z\ dust features in our images, which have
properties similar to those identified in \ngc 891 (Howk \& Savage
1999c; \pII) correspond to high-\z\ CO emission in the maps of
\garciaburillo\ \etal\ The direct observation of molecular emission
associated with the extraplanar dust structures in \ngc 4013 supports
our suggestion in \pII\ that the high-\z\ absorbing dust structures
observed in several edge-on galaxies represent a dense, possibly
molecular phase of the thick disk ISM.  \garciaburillo\ \etal\ (1992)
and Sofue \& Nakai (1993) have also given evidence for a CO-emitting
component of the thick disk of
\ngc 891, though at much lower resolution.

Several candidate \HII\ regions can be identified in our \Ha\ images
of \ngc 891 at heights $z \sim 0.6$ to 2.0 kpc from the midplane (Howk
\& Savage 1999c; Howk \& Savage, in preparation).  Associated with
these candidate \HII\ regions we find faint continuum sources ($B \sim
22$ to 23 mag).  If spectroscopically confirmed, the presence of \HII\
regions at such high-\z\ suggests the underlying stars were formed in
the thick interstellar disk of \ngc 891.  This would require the
existence of molecular clouds far from the midplane, and hence lend
implicit support to our conclusion that a dense phase of the ISM is
present at high-\z, visible as the absorbing dust structures seen in
our images.  These nebulae will also be a useful probe of the physical
conditions of the thick disk ISM in \ngc 891.  For example, our WIYN
images constrain the radii of these nebulae to be $r\la20$ pc.  If
ionized by B0 or earlier stars, this implies an ambient electron
density of at least $n_e \ga 1$ \percc\ (Osterbrock 1989).  Thus the
density of at least one phase of the interstellar thick disk of \ngc
891 is quite dense, even at rather large heights above the plane
($z\sim2$ kpc).  Density estimates for the DIG at these heights are
factors of $3-10$ smaller than the lower density limit implied the
\HII\ regions, further suggesting large variations in the density of the
interstellar thick disk.

We note that the dust filaments seen in our images likely cannot be
caused by a gaseous warp in this galaxy, and they are unlikely to be
associated with a flare in the outer galaxy.  \ngc 891 shows no
evidence for a warp in its \HI\ distribution (Swaters \etal\ 1997;
Sancisi \& Allen 1979), and the relative numbers of the dust features
on either side of the midplane are very similar (\pI), suggesting
these features cannot come from a warp along the line of sight.  While
a flared gaseous layer cannot be ruled out, it seems unlikely the
observed dust features are associated with a flare.  At least {\em
some} of the features are clearly associated with regions of vigorous
star formation in the disk.  The detection of CO from similar
extraplanar dust structures in \ngc 4013 by \garciaburillo\ \etal\
(1999) gives some information on the velocities of these structures.
In particular, one such dust feature lies immediately above the
dynamical center of the galaxy, but is moving at $-104$ \kms\ relative
to the systemic velocity of the galaxy.  This velocity is clearly
incompatable with the location of the structure in a smoothly rotating
flared gas layer.  Several other features can be identified with
velocities inconsistent with their presence in a galactic flare.  We
do not believe a flared or warped gas layer is causing the dust
structures we see in our images.

We believe a self-consistent picture of the high-\z\ ISM in
\ngc 891 can be drawn in which the thick disk ISM is made
up of a multiphase medium, including a cold, dense phase, a warm
(neutral and ionized) phase, and a hot phase.  The estimated masses of
the warm neutral, warm ionized, and cold, dense phases of the ISM
above $z\sim400$ pc are all of order a few$\times\e{8}$ \msun.  The
high-\z\ phases of the ISM together represent $\sim10\%$ of the total
mass of the ISM in \ngc 891 (Swaters \etal\ 1997; Dettmar 1990;
\garciaburillo\ \etal\ 1992; \pI).  The mass of the hot ISM at
high-\z\ is estimated to be $\sim4\times\e{7}$
\msun\ (Bregman \& Houck 1997).  The presence of the dense phase, 
in particular, requires a relatively large pressure to large heights
above the midplane.  The available data on the hot and warm ionized
media are consistent with the required pressures in the thick disk of
\ngc 891.

Such high pressure in this transition region between the thin disk and
the more extended halo of the galaxy is required to support the weight
of the higher-\z\ gas.  However, the origin of this high-\z\ gas, and
hence the origin of the high pressure confining the dense clouds, is
likely ultimately tied to star formation in the disk of the galaxy.
And though the existence of interstellar material in the thick disks
and halos of galaxies was once thought to be a relatively rare
phenomenon, our previous work (\pII) implies the transfer of matter
from the thin interstellar disks to the more extended halos may be a
common property of spiral galaxies.  Furthermore, our images imply
that this transfer does not destroy the dust grains.  This is
important since it can therefore be argued that the more extended
halos of galaxies may not be completely devoid of dust (see, e.g.,
Zaritsky 1994).


\section{Summary}
\label{sec:summary}

We have presented deep, high-resolution ground based images of the
edge-on spiral \ngc 891 obtained with the WIYN 3.5-m telescope.  These
images were used to study the extensive distribution of extraplanar
dust and ionized gas in this galaxy, building on the work presented in
\pI.  The major results of our work are as follows.

\begin{enumerate}

\item Our deep broadband images show an extensive web of
inhomogeneously distributed (clumped) dust far from the midplane of
\ngc 891.  Many of these structures were visible in the images
presented in \pI.  These new images presented here show individual
absorbing dust structures to heights $z\ga2.5$ kpc from the midplane.
The number and apparent extinctions of these dust structures decreases
significantly beyond $z\sim1.7 - 2.0$ kpc.  Structures like those seen
at $z\sim 0.5 - 1.0$ kpc would be easily observable to heights
$z\sim3$ kpc if they existed at such heights.

\item We identify a small subset of the high-\z\ absorbing structures
and derive their physical properties (see Tables \ref{table:observed}
and \ref{table:derived}).  These structures are characterized by sizes
of order 50-300 pc, with optical extinctions exceeding $A_V \sim 0.2 -
0.8$ mag.  Assuming Galactic gas to dust relationships the extinction
values imply the column density of associated gas in these structures
is $N_{\rm H} \ga \e{21}$ \column.  The implied mass of the associated
gas is $M \ga \e{5}$ \msun\ per cloud, and the gravitational potential
energies relative to the midplane are large, $\Omega \ga \e{52}$ ergs.
The masses and average extinction values derived for these clouds are
reminiscent of the Galactic GMCs, and in general we conclude that the
dust structures are tracing the dense cold neutral phase of the
high-\z\ ISM.

\item A few of the more spectacular absorbing dust structures are
discussed in some detail.  These include a cometary structure of width
$\sim140$ pc and length $\sim500$ pc at a height $z\sim1.3$ kpc from
the midplane; a 600-pc diameter supershell centered at $z\sim0.8$ kpc;
and a filament that may stretch from $z\sim0.2$ to $\sim2.0$ kpc.

\item Our deep, high-resolution \Ha\ emission line images recover the
high-\z\ DIG studied in earlier works (e.g., Rand \etal\ 1990; Dettmar
1990; Pildis \etal\ 1994).  These data show a filamentary component of
extraplanar ionized gas as well as an apparently more diffuse
component.  In general the DIG emission is much more smoothly
distributed than the interstellar material traced by the dust
absorption.  The northeast section of \ngc 891 shows significantly
more emission, both in and out of the plane, than the opposite side of
the disk.

\item Though filaments are indeed present in the distribution of the DIG, 
they are much less numerous and prominent than the filamentary
dust-bearing clouds.  The excess emission in the filaments is between
30\% and 100\% the brightness of the background DIG emission.  At low
\z\ ($\la 1$ kpc) much of the structure seen in the \Ha\ images is the
result of absorption by the same dust-bearing clouds visible in our
broadband images.  We identify several ionized, filamentary structures
and summarize their rough properties.  Our catalog of structures
includes a spectacular supershell of diameter $\sim650$ pc that is
centered $\sim1500$ pc above the plane of \ngc 891.

\item {\em A direct comparison of the DIG morphology with that of the 
complicated extraplanar dust distribution shows little correspondence
between the two.}  There are many regions where the \Ha\ is extincted
by the dust structures in our images, but very few that show a direct
physical relationship between these two tracers of high-\z\ material.
In the northeast section of the disk, a region showing evidence for
very vigorous star formation, there are a few structures that show a
loose association between the thick disk DIG and dust.  The
morphologies of these structures strongly suggest they are connected
to the energetic effects of star formation in the underlying disk.

\item The high-\z\ extinction structures seen in our
images represent a dense, likely cold, {\em phase} of the high-\z\
multiphase ISM.  This conclusion is based on the lower limits to the
dusty cloud extinctions, the mass and density estimates, and the clear
separation between the high-\z\ medium traced by the dust structures
and that traced by emission from the DIG.  Secondary support for this
conclusion is provided by the possible existence of star formation,
evidenced by several \HII\ regions visible in our \Ha\ data, at large
distances from the plane; and the recent detection of CO emission from
similar dust structures in the galaxy \ngc 4013 by \garciaburillo\
\etal\ (1999).  The thermal pressure of the hot and warm ionized gas  
is sufficient to maintain a stable two-phase neutral medium
(warm+cold) at large distances from the plane (Wolfire \etal\ 1995a),
the densest portions of which we are seeing in our images.  The
estimated masses of the cold dense, warm neutral, and warm ionized
phases at $z > 400$ pc are comparable and of the order a few$\times
\e{8}$ \msun.

\item The presence of \HII\ regions, their underlying young stellar 
populations, and the natal molecular clouds from which these stars
have formed in the thick disk of \ngc 891 provides an opportunity to
study the physics of star formation in an extreme environment.  Our
broadband images give a direct visual means of studying the dense
high-\z\ clouds in this galaxy with superb spatial resolution (for
studies of this phase of the ISM).

\end{enumerate}

\acknowledgements

It is a pleasure to thank the many people who have made the WIYN
telescope a reality.  We extend special thanks to those who maintain
and operate observatory.  JCH recognizes support from a NASA Graduate
Student Researcher Fellowship under grant number NGT-5-50121.

This research has made use of the NASA/IPAC Extragalactic Database
(NED) which is operated by the Jet Propulsion Laboratory, California
Institute of Technology, under contract with the National Aeronautics
and Space Administration.


\pagebreak

\newcommand{\Haon}{H$\alpha$}
\newcommand{\Hahi}{Off 2}
\newcommand{\Halo}{Off 1}

\begin{planotable}{clcc}
\tablewidth{0pc}
\tablecolumns{4}
\tablecaption{Log of Observations\label{table:log}}
\tablehead{
\colhead{Filter} &
\colhead{Date}   & 
\colhead{Exposure} &
\colhead{Seeing} \nl
\colhead{} &\colhead{[UT]} &\colhead{[sec]} &
\colhead{[arcsec]} }
\startdata
\multicolumn{4}{c}{Broadband Observations} \nl
\cline{1-4}
 B & 1997 August 30   &  1200 &  0.8  \nl 
 B & 1997 August 30   &   900 &  0.6  \nl 
 B & 1997 August 30   &   900 &  0.7  \nl 
 B & 1997 August 31   &  1200 &  0.8  \nl 
 B & 1997 August 31   &  1200 &  0.8  \nl 
 V & 1996 December 04  &  120  & 0.7  \nl 
 V & 1997 August 30    &  1200 & 0.7  \nl 
 V & 1997 August 30    &  1200 & 0.8  \nl 
 V & 1997 August 31    &  900  & 0.6  \nl 
 I & 1997 August 30   &   840 & 0.7  \nl 
 I & 1997 August 30   &  1500 & 0.7  \nl 
 I & 1997 August 31   &  1200 & 0.7  \nl 
 I & 1997 November 28 &   820 & 0.8  \nl 
\cline{1-4}
\multicolumn{4}{c}{Narrowband Observations} \nl
\cline{1-4}
\Haon  &  1997 August 31    & 1200  &  0.7   \nl 
\Haon  &  1997 August 31    &  300  &  0.8   \nl 
\Haon  &  1997 November 26  & 1800  &  0.6   \nl 
\Haon  &  1997 November 26  & 1800  &  0.6   \nl 
\Haon  &  1997 November 26  & 1800  &  0.6   \nl 
\Haon  &  1997 November 26  & 1800  &  0.7   \nl 
\Haon  &  1997 November 29  & 1200  &  0.6   \nl 
\Haon  &  1997 November 29  & 1800  &  0.8   \nl 
\Haon  &  1997 November 29  & 1200  &  0.8   \nl 
\Haon  &  1997 November 29  & 1200  &  0.7   \nl 
\Halo  &  1997 November 26  & 1800  &  0.6   \nl 
\Halo  &  1997 November 26  & 1800  &  0.6   \nl 
\Halo  &  1997 November 29  & 1200  &  0.8   \nl 
\Hahi  &  1997 August 30    & 1200  &  0.9   \nl 
\Hahi  &  1997 August 31    &  900  &  0.7   \nl 
\enddata
\end{planotable}

\pagebreak

\begin{planotable}{ccc}
\tablewidth{0pc}
\tablecolumns{3}
\tablecaption{Final Dataset\label{table:finalimages}}
\tablehead{
\colhead{Filter} &
\colhead{Exposure} &
\colhead{Seeing} \nl
\colhead{} &
\colhead{[sec]} &
\colhead{[arcsec]} }
\startdata
B	& 5400 & 0.77 \nl
V 	& 3420 & 0.80 \nl
I	& 4360 & 0.81 \nl
\Haon
	& 14100 & 0.77\tablenotemark{a} \nl
\Haon\ Off 
	& 6900 &  0.90 \nl
\enddata
\tablenotetext{a}{Though the on-band \Haon\ data have a resolution of
	0\farcs77, we have smoothed these data to match the 0\farcs9
	resolution of the off-band data to produce the final \Haon\
	image.} 
\end{planotable}

\pagebreak

\begin{planotable}{lcc}
\tablewidth{0pc}
\tablecolumns{3}
\tablecaption{Narrow Band Filter Characteristics\label{table:narrowfilters}}
\tablehead{
\colhead{Filter} &
\colhead{$\lambda_{cen}$} &
\colhead{FWHM} \nl
\colhead{} &
\colhead{[\AA]} &
\colhead{[\AA]} }
\startdata
\Haon (W015)  & 6570 & 73 \nl
\Halo (KP809) & 6488 & 67 \nl
\Hahi (W016)  & 6618 & 72 \nl
\enddata
\end{planotable}

\pagebreak

\begin{landscape}
\begin{deluxetable}{lcccccccl}
\tablenum{4}
\tablewidth{0pc}
\tablecolumns{9}
\tablecaption{Observed Properties of Individual High-$z$ Dust
Features\label{table:observed}}
\tablehead{
\colhead{ID\tablenotemark{a}} & 
\colhead{R.A.} & \colhead{Dec.} & 
\colhead{$|z|$\tablenotemark{b}} & 
\colhead{Dimensions} &
\colhead{$a_{B}$\tablenotemark{c}}  & 
\colhead{$a_{V}$\tablenotemark{c}}  & 
\colhead{$a_{I}$\tablenotemark{c}}  &
\colhead{Morphology} \nl
\colhead{[NGC 0891:D]}  &  
\colhead{[J2000]} & \colhead{[J2000]} &
\colhead{[pc]}   & \colhead{[pc$\times$pc]} &
\colhead{[mag.]} &  
\colhead{[mag.]} &
\colhead{[mag.]} & 
\colhead{} 
}
\startdata
$-119-019$ 
	& 02 22 30.8  & +42 18 55	    
	& \phn870  & $270\times75$\phn  & 0.37  & 0.22  &  0.15
	& Flaring Cloud \nl
$-107-020$ 
	& 02 22 31.2  & +42 19 06	    
	&  \phn880  & $200\times100$  & 0.50  & 0.35 & 0.19
	& Irr. Cloud  \nl
$-089+039$ 
	& 02 22 26.9  & +42 19 45	    
	& 1750  & $300\times160$  & 0.28  & 0.25 & 0.19
	& Irr. Cloud  \nl  
$-044+033$\tablenotemark{d}
        & 02 22 29.1 & $+42$ 20 23
        & 1450 & $100\times50$\phn & 0.77 & 0.56 & 0.30  
	& Elongated Cloud \nl  
$-012-030$\tablenotemark{e}
        & 02 22 35.2 &  $+42$ 20 29 
        & 1300 & $140\times60$\tablenotemark{f}\phn  
		& 0.36 & 0.32 & 0.17   
	&  Cometary Cloud \nl   
$-010+037$ 
	& 02 22 29.8  & +42 20 57     
	&  1660 & $240\times120$  & 0.32  & 0.21 &  0.16
	& Arc \nl 	
$+007-032$ 
	& 02 22 36.2  & +42 20 47	    
	& 1430  & $110\times90$\tablenotemark{f}\phn   
		& 0.27  & 0.24  &  0.13
	& Cometary Cloud \nl
$+017+016$\tablenotemark{g}
	& 02 22 32.5  & +42 21 14
	& \phn730 &  $350\times75$\phn & 1.02 & 0.83 & 0.49
	&  Column \nl 
$+033+043$
	& 02 22 30.8 & +42 21 40 
	& 1950 & $200\times55$\phn & \nodata\tablenotemark{h} 
	& \nodata\tablenotemark{g} & \nodata\tablenotemark{h}
	& Irr. Cloud \nl 
\tablebreak
$+105-017$\tablenotemark{i}
	& 02 22 39.0 & +42 22 34 
	& \phn750 & $700\times290$  & 0.50 & 0.40 & 0.27 
	&  Vertical Cone \nl 
$+122-016$\tablenotemark{j}
	& 02 22 38.9 & +42 22 39
	& \phn730  & $270\times60$\phn  & 0.63 & 0.56 &  0.31
	&  Shell Wall \nl
$+134-019$\tablenotemark{k}
	 & 02 22 39.6 & +42 22 50 
	& \phn845 & $400\times50$\phn  & 0.40 & 0.35 & 0.23
	& Shell Wall \nl 
\enddata
\tablenotetext{a}{ Identification of the feature using the
nomenclature NGC 0891:D $\pm$XXX$\pm$ZZZ (see text).  We only list
$\pm$XXX$\pm$ZZZ in the main body of the table.}
\tablenotetext{b}{ Projected height above the midplane, or
the limit to which very extended features can be traced.}
\tablenotetext{c}{ Average apparent extinction for the BVI wavebands
in magnitudes, as defined in the text.}
\tablenotetext{d}{Identified as feature 2 in Paper~I.}
\tablenotetext{e}{Identified as feature 7 in Paper~I.}
\tablenotetext{f}{The size of the head of these cometary structures
	is listed.  The overall extent of each, including the
	potential tale material, is $\sim500$ pc for $-012-030$ and
	$\sim200$ pc for $+007-032$.}
\tablenotetext{g}{Identified as feature 4 in Paper~I.}
\tablenotetext{h}{This feature lies in front of a background
	elliptical galaxy, making its properties difficult to
	measure.}
\tablenotetext{i}{Identified as feature 10 in Paper~I.}
\tablenotetext{j}{Identified as feature 11a in Paper~I.}
\tablenotetext{k}{Identified as feature 11b in Paper~I.}
\end{deluxetable}
\end{landscape}


\pagebreak

\begin{deluxetable}{ccccccc}
\tablenum{5}
\tablewidth{0pc}
\tablecolumns{7}
\tablecaption{Derived Properties of Individual High-$z$ Dust
Features\label{table:derived} }
\tablehead{
\colhead{ID\tablenotemark{a}} & 
\multicolumn{3}{c}{$a_{V}(x)$ [mag.]\tablenotemark{b}} &
\colhead{$N_{\rm H}$\tablenotemark{c}}  & 
\colhead{Mass\tablenotemark{d}} & \colhead{Energy\tablenotemark{e}}\nl
\cline{2-4}
\colhead{[NGC 0891:D]} &
\colhead{$x = 0$}  & \colhead{$x = 0.25$}  & \colhead{$x = 0.5$}  & 
\colhead{[cm$^{-2}$]} &  
\colhead{[M$_\odot$]} & \colhead{[ergs s$^{-1}$]} }
\startdata
$-119-016$ & 0.22 & 0.30 & 0.50 
	& $>4\times\e{20}$ & $>1\times\e{5}$ & $>9\times\e{51}$ \nl
$-107-017$ & 0.35 & 0.50 & 0.87 
	& $>7\times\e{20}$ & $>9\times\e{4}$ & $>6\times\e{51}$ \nl
$-089+039$ & 0.25 & 0.35 & 0.58 
	& $>5\times\e{20}$ & $>3\times\e{5}$ & $>6\times\e{52}$ \nl
$-044+032$ & 0.56 & 0.84 & 1.79 
	& $>1\times\e{21}$ & $>1\times\e{5}$  & $>2\times\e{52}$ \nl
$-012-030$&  0.32 & 0.45 & 0.78
	& $>6\times\e{20}$ & $>1\times\e{5}$ & $>2\times\e{52}$  \nl
$-010+037$ & 0.21 & 0.29 & 0.47  
	& $>4\times\e{20}$ & $>2\times\e{5}$ & $>3\times\e{52}$ \nl
$+007-032$ & 0.24 & 0.33 & 0.55 
	& $>5\times\e{20}$ & $>4\times\e{4}$ & $>6\times\e{51}$ \nl
$+017+016$ & 0.83 & 1.35 & \nodata
	& $>2\times\e{21}$ & $>6\times\e{5}$ & $>3\times\e{52}$ \nl
$+105-017$ & 0.40 & 0.57 & 1.04 
	& $>8\times\e{20}$ & $>9\times\e{5}$ & $>4\times\e{52}$ \nl
$+122-016$ & 0.56 & 0.84 & 1.78 
	& $>1\times\e{21}$ & $>5\times\e{5}$ & $>3\times\e{52}$ \nl
$+134-019$ & 0.35 & 0.50 & 0.87
	& $>7\times\e{20}$ & $>3\times\e{5}$ & $>2\times\e{52}$ \nl
\enddata
\tablenotetext{a}{Identification of the feature using the
nomenclature NGC 0891:D $\pm$XXX$\pm$ZZZ, where XXX is the
distance in arcsec from the optical center of the galaxy traced along
the major axis, and ZZZ is the distance in arcsec from the optical
center of the galaxy traced along the minor axis (see text).  We only
list $\pm$XXX$\pm$ZZZ in the main body of the table.}
\tablenotetext{b}{Lower limits to the V-band extinction assuming a
fraction $x$ of the stellar light resides in front of the dust
feature, given for $x=0,$ 0.25, and 0.5.  The listed values $a_V(x)$
are lower limits to the true extinction, $A_V$, due to our neglect of
scattering into the line of sight.}
\tablenotetext{c}{Approximate lower limit to the column
density of material assuming $N_{\rm H} > 1.9\times~\e{20} a_V (x=0)$
cm$^{-2}$ (see text).}
\tablenotetext{d}{Approximate mass based upon the
estimated column density and projected area.  Includes a factor of
1.37 correction for He.}
\tablenotetext{e}{Rough estimates for the potential energy of each
structure relative to the midplane using the projected distance from
the midplane and the estimated mass (see text).}
\end{deluxetable}

\pagebreak

\begin{deluxetable}{lcccc}

\tablewidth{0pc}
\tablecolumns{9}
\tablecaption{Observed Properties of Individual DIG 
Filaments\label{table:filaments}}
\tablehead{
\colhead{ID\tablenotemark{a}} & 
\colhead{R.A.} & \colhead{Dec.} & 
\colhead{$|z|$\tablenotemark{b}} & 
\colhead{Dimensions} \nl
\colhead{[NGC 0891:DIG]}  &  
\colhead{[J2000]} & \colhead{[J2000]} &
\colhead{[pc]}   & \colhead{[pc$\times$pc]} }
\startdata
$+026+035$  & 02 22 31  &  +42 21 30 &    1570 & $750\times150$ \nl
$+089-034$\tablenotemark{c}
            & 02 22 39  &  +42 22 02 &    1510 & $600\times180$ \nl
$+111+024$  & 02 22 35  &  +42 22 45 &    1100 & $700\times130$ \nl
$+123-018$\tablenotemark{d}
            & 02 22 39  &  +42 22 40 & \phn820 & $500\times100$ \nl
$+128+015$\tablenotemark{e}
            & 02 22 36  &  +42 22 58 & \phn660 & $400\times60$ \nl
$+133-011$\tablenotemark{d}
            & 02 22 39  &  +42 22 52 & \phn510 & $270\times90$ \nl 
$+149-022$  & 02 22 40  &  +42 23 02 &    1010 & $700\times50$ \nl
\enddata
\tablenotetext{a}{ Identification of the H$\alpha$-emitting filament 
using the nomenclature NGC 0891:DIG $\pm$XXX$\pm$ZZZ (see text).  We
only list $\pm$XXX$\pm$ZZZ in the main body of the table.}
\tablenotetext{b}{ Projected height above the midplane, or
the limit to which very extended features can be traced.}
\tablenotetext{c}{ This structure is in fact a large shell.  The 
height and position listed in the table are appropriate for the center
of the shell.  The dimensions are appropriate for the thickness and
rough vertical length of the shell walls.  The shell itself is $\sim650$
pc in diameter.}
%
%
\tablenotetext{d}{ This filament is associated with the shell formed by the 
dust structures NGC 0891:D $+122-016$ and $+134-019$.  Together they make up one
of the ``supershells'' discussed by Pildis \etal\ 1994.}
\tablenotetext{e}{ This structure is part of one  of the ``supershells'' discussed 
by Pildis \etal\ 1994.  In our higher resolution image it is not clear
that this structure truly represents a supershell.}
\end{deluxetable}


\pagebreak
\clearpage

\pagebreak

\clearpage

\end{document}